\documentstyle[pra,eqsecnum,aps]{revtex}

\def\displayfrac#1#2{\frac{\displaystyle #1}{\displaystyle #2}}

\begin{document}
\title{Generalised Coherent States and the Diagonal Representation
for Operators}

\author{N. Mukunda\thanks{Permanent Address: Centre for Theoretical
Studies and Department of Physics,
 Indian Institute of Science, {Bangalore~560~012,} India, and  
Jawaharlal Nehru Centre for Advanced Scientific Research,
 Jakkur, Bangalore~560~064, India}
\thanks{email: nmukunda@cts.iisc.ernet.in}}
\address{Dipartimento di Scienze Fisiche, Universita di Napoli
``Federico II''\\
Mostra d'Oltremare, Pad. 19--80125, Napoli, Italy\\
and\\
Dipartimento di Fisica dell Universita di Bologna\\
Viale C.Berti Pichat, 8 I--40127, Bologna, Italy}
\author{ Arvind\thanks{email:arvind\@physics.iisc.ernet.in}} 
\address{Department of Physics, Guru Nanak Dev University, Amritsar 143005, India}
\author{S. Chaturvedi\thanks{e-mail: scsp\@uohyd.ernet.in}} 
\address{ School of Physics, University of Hyderabad, Hyderabad 500 046,
India}
\author{ R.Simon\thanks{email: simon\@imsc.ernet.in}}
\address{The Institute of 
 Mathematical Sciences, C. I. T. Campus, Chennai 600 113, India}
 \date{\today}
 \maketitle

\begin{abstract}
We consider the problem of existence of the diagonal representation for 
operators  in the space of a family of generalized coherent states 
associated with an unitary irreducible representation of a (compact) Lie group.
We show that necessary and sufficient conditions for the possibility of such a
representation can be obtained by combining Clebsch-Gordan theory and the
reciprocity theorems associated with induced unitary group representation. 
 Applications to several examples involving $SU(2),$ $SU(3),$ and the
Heisenberg-Weyl group are presented, showing that there are simple examples
 of generalized coherent states which do not meet these conditions. Our results
are relevant for phase-space description of quantum mechanics and quantum state
reconstruction problems. 
\end{abstract}
\pacs{PACS: abns   }


\section{Introduction}

There is a long history of attempts to express the basic
structure of quantum mechanics, both kinematics and
dynamics, in the $c$-number phase space language of
classical mechanics.  The first major step in this
direction was taken by Wigner\cite{Wigner} very early in the 
development of quantum mechanics, during a study of
quantum corrections to classical statistical mechanics.
This led to the definition of a real phase space
distribution\cite{Hillery}- now called the Wigner distribution - 
faithfully representing any pure or mixed state of a 
quantum system whose kinematics is governed by
Heisenberg commutation relations for any number of
Cartesian degrees of freedom.  It was soon realized that
this construction is dual to a rule proposed earlier
by Weyl\cite{Weyl} to map classical dynamical variables onto
quantum mechanical operators in an unambiguous way,
in the sense that the expectation value of any
quantum operator in any quantum state can be rewritten
in a completely $c$-number form on the corresponding
classical phase space.

The general possibilities of expressing quantum mechanical 
operators in classical $c$-number forms were later examined
by Dirac\cite{Dirac} while developing the analogies between classical
and quantum mechanics.  The specific case of the Weyl-Wigner 
correspondence was carried further in important work by 
Groenewold and by Moyal\cite{Groenewold}.

Inspired by the needs of quantum optics, the general problem
of setting up different classical variable - quantum operator 
correspondences has received enormous attention\cite{Agwocg}.  It has 
thus been appreciated that the Weyl-Wigner choice is just 
one of many possibilities, two other important ones being
(in the language of photon annihilation and creation 
operators) the normal ordering \cite{GS}and the antinormal 
ordering\cite{Husimi} choices.  In this same context, the coherent states of 
the harmonic oscillator with their remarkable properties have
played a crucial role.  At the state vector level these states --
right eigenstates of the annihilation operator with  complex
eigenvalues -- form an overcomplete system.  At the operator
level this same overcompleteness can be exploited to prove
that any operator can be expressed as an integral over 
projections onto coherent states\cite{GS}, though in general the
accompanying coefficient `function' can be a very singular 
distribution\cite{KSMW}.  This operator diagonal coherent state 
representation is dual to the normal ordering rule in the
same sense as the Wigner and Weyl rules are dual to one 
another.

The idea of coherent states has been extended, in two slightly different 
ways, by Klauder\cite{Klauder}
and by Perelomov\cite{Perelomov}, to the notion of generalised coherent 
states. A very interesting case is the family of coherent states in the
context of an unitary irreducible 
representation (UIR) of any Lie group $G$ on a Hilbert
space ${\cal H}$. 
In particular, the generalised coherent 
states associated with the group $SU(2)$ -- the atomic coherent states -- 
have been investigated, in detail, in the literature\cite{Arrechi}. 
The (over) completeness property of the generalised coherent states, at the 
vector space level, usually (but not always)  follows as a direct
consequence
of Schur's lemma.  This depends on whether the group representation
possesses the square integrability property or not. On the other hand, the
concept of the Wigner distribution has been more or less directly 
generalised to other kinds of quantum kinematics, notably the 
case of spin described by the appropriate representations of
$SU(2)$\cite{Agarwal}.

The previous remarks lead to the natural question whether
in the case of a system of generalised coherent states too
the diagonal representation for operators continues to exist, 
as a counterpart to the (over) completeness at the vector level.
In our opinion this question has hitherto not received the 
direct attention that it deserves.  However, we should invite the reader's 
attention to some very insightful remarks by Klauder and Skagerstam on this 
question\cite{klauderreprints}. 

The main aim of the present
paper is to examine this question and to develop necessary
and sufficient conditions which will ensure that all operators
on the (relevant) Hilbert space can indeed be expanded in
terms of projections onto generalised coherent states.  For 
definiteness we deal with the situation where the group $G$
is compact, so that its chosen UIR acts on a Hilbert space
${\cal H}$ of finite dimension.  (However in some of 
the examples we formally extend our methods to certain
non compact $G$).  The important tools in our analysis 
are certain well-known reciprocity theorems when one 
examines an induced representation of $G$\cite{Mackey} arising from 
some UIR of a subgroup $H\subset G$ and asks for the
occurrence and multiplicity of various UIR's of $G$ itself; 
and the structure of the Clebsch-Gordan series and
coefficients for direct products of UIR's of $G$, in a form 
adapted to $H$.  We will show that while the necessary and 
sufficient conditions mentioned above are met in certain cases
of $SU(2)$ and the Heisenberg-Weyl (H-W) group, there are
quite simple examples in the cases of $SU(2)$ and $SU(3)$ where they
are not satisfied.  This will attest to the necessity
and significance of the conditions that we develop.

In important recent work Brif and Mann\cite{Brim} have approached
this family of questions in a general way, exploiting the
tools of harmonic analysis on coset spaces.  However, 
their results do not include a complete set of necessary
and sufficient conditions for the possibility of 
diagonal representations of operators, nor is there an
indication that there are fairly simple situations where
such representations are not possible.

The contents of this paper are organised as follows.  In
Section 2 we set up the basic notations and definitions
of generalised coherent states within a UIR of a general 
compact Lie group $G$, the two associated stability groups
and coset spaces, and carry out the harmonic analysis at
the vector level.  The two distinct kinds of relationships
between the stability groups are also carefully defined.
Section 3 discusses the detailed properties of the
projection operators onto the generalised coherent
states, and performs the corresponding harmonic analysis.
Using these and other results pertaining to the Clebsch-
Gordan problem, we are able to obtain explicit necessary
and sufficient conditions for existence of the diagonal 
representation in any given situation.  In Section 4 we
consider applications to both $SU(2)$ and $SU(3)$, taking
three and two examples respectively.  The aim is to show
how to check our conditions in practical cases, and to 
exhibit some simple situations where the diagonal
representation exists, and other equally simple ones 
where it does not.  Section 5 analyses the Heisenberg-Weyl
group in a heuristic way, to display how our conditions
work and lead to expected results.  Section 6 contains
concluding remarks.  Appendices A and B gather material
on general Clebsch-Gordan series and coefficients, unit
tensor operators, induced representation theory and the
reciprocity theorem.

\section{Harmonic analysis on coset spaces -- the vector level}
\setcounter{equation}{0}

Let $G$ be an $n$ dimensional compact Lie group.  As described 
in Appendix $A$, we denote the various UIR's (upto equivalence)
of $G$ by a symbol $J$; within a UIR we denote a complete set of
orthonormal basis labels (magnetic quantum numbers) by $M$.  Both
$J$ and $M$ stand in general for sets of several independent indices.
Certain specific choices of the latter will be indicated later.

Let the Hilbert space ${\cal H}_{J_{0}}$ carry the $N_{J_{0}}$ 
dimensional UIR ${\cal D}^{(J_{0})}(\cdot)$ of $G$.  Choose
and keep fixed some fiducial unit vector $\psi_0\,\in\,
{\cal H}_{J_{0}}$.  The orbit of $\psi_0$ is the collection 
of vectors - generalised coherent states - $\psi(g)\,\in\,
{\cal H}_{J_{0}}$ obtained by acting on $\psi_0$ with all
$g\,\in\,G$:
     \begin{eqnarray}
     \vartheta(\psi_0) = \left\{\psi(g) =
     {\cal D}^{(J_{0})}(g) \psi_0\;\bigg|\;
     g\,\in\, G\right\} \subset
     {\cal H}_{J_{0}}\,.
     \end{eqnarray}

\noindent
Similarly, if $\rho_0 = \psi_0 \psi_{0}^{\dag}$ is the pure state
density matrix corresponding to $\psi_0$, its orbit in the
space of all density matrices is
     \begin{eqnarray}
     \vartheta(\rho_0) =\left\{\rho(g) ={\cal D}^{(J_{0})}
     (g) \rho_0 \;{\cal D}^{(J_{0})} (g)^{\dag} =
     \psi(g) \psi(g)^{\dag} \;\bigg|\;  g  \,\in\,G\right\}\,.
     \end{eqnarray}

\noindent
Two important subgroups $H_0, H$ in $G$ are now defined:
     \begin{eqnarray}
     H_0 &=& \left\{g\,\in\,G \;\bigg|\; {\cal D}^{(J_{0})}(g)
     \psi_0 =\psi_0 \right\} \subset G ,\nonumber\\
     H &=&\left\{g\,\in\,G\;\bigg|\; {\cal D}^{(J_{0})}(g)
     \psi_0 = (\mbox{phase}) \;\psi_0\right\}\subset G\,.
     \end{eqnarray}

\noindent
The dependences of $H_0, H$ on $\psi_0$ are left implicit.  
The subgroup $H_0$ is the stability group of $\psi_0$ in
the strict sense, while $H$ is the stability group of 
$\psi_0$ upto phase factors.  On the other hand, $H$
is the stability group of $\rho_0$ in the strict sense:
     \begin{eqnarray}
     H = \{g\,\in\,G \;| \;\rho(g) = \rho_0\} \subset G\,.
     \end{eqnarray}

\noindent
By standard arguments one has the identifications of the
two orbits with corresponding coset spaces of $G$.
     \begin{eqnarray}
     \vartheta(\psi_0) &\simeq & G/H_0 = \Sigma_0 \;,\nonumber\\
     \vartheta(\rho_0) &\simeq & G/H = \Sigma\;.
     \end{eqnarray}

\noindent
For definiteness we always take coset spaces to be made up
of right cosets $gH_0, gH$ in the two cases.

It is evident that $H_0$ is an invariant subgroup of $H$, and we 
can distinguish two qualitatively different situations depending 
on the nature of the quotient $H/H_0$ :
     \begin{eqnarray}
     \mbox{\bf Case A:}\;\;\;\;H/H_0 &=& \mbox{trivial or
discrete}\,,\nonumber\\
     \mbox{\bf Case B:}\;\;\;\; H/H_0 &=& U(1)\,.
     \end{eqnarray}

\noindent
These two possibilities can be pictured as follows:
There is an obvious and natural projection map 
$\pi : \vartheta(\psi_0)\rightarrow \vartheta(\rho_0)$
or $\pi : \Sigma_0\rightarrow \Sigma$.  (Since $H_0$
is a subgroup of $H$, every $H_0$-coset lies within 
some $H$-coset).  With respect to this projection map,
in Case A for each $\rho \,\in\,\vartheta(\rho_0)$,
there is just one or a discrete set of vectors $\psi\,
\in\,\pi^{-1}(\rho)\subset\vartheta(\psi_0)$; 
while in Case B $\pi^{-1}(\rho)$ consists of all 
vectors $\{e^{i\alpha}\psi\}$ for some fixed $\psi$
and $0\leq\alpha < 2\pi$.  Stated in yet another manner : 
in Case A with the help of action by elements in $G$ 
the phase of $\psi_0$ (and so of any $\psi(g)$) can be
altered in only a discrete set of ways or not at all; 
and in Case B these phases can be altered in a continuous 
manner, so that each $\pi^{-1}(\rho)$ contains a
``U(1)- worth of vectors''.

We now wish to exploit the results of harmonic analysis
arising from the natural UR's of $G$ acting on square 
integrable functions on the two coset spaces $\Sigma_0, 
\Sigma$ in order to extract the $G$ representation 
contents of $\psi(g), \rho(g)$ respectively.  The key 
point is that while both $\psi(g)$ and $\rho(g)$ have 
already known dependences on $g$, since they are 
obtained from $\psi_0$ and $\rho_0$ respectively by 
actions via the given UIR ${\cal D}^{(J_{0})}$ of $G$ 
(and in particular $\psi(g)$ for different $g$ may 
not be orthogonal, $\rho(g)$ for different $g$ may 
not be  trace orthogonal), they are linear quantities.
Namely each of them belongs to a corresponding linear
space.  Therefore natural complete orthonormal sets
of functions on $\Sigma_0, \Sigma$ can be profitably
used to project out the irreducible `Fourier' 
components of $\psi(g), \rho(g)$ respectively with
well defined irreducible behaviour under $G$, and
then to resynthesize them.  In the remainder of this 
Section we look at the case of $\psi(g)$, ie., we
consider the situation at the vector level.  In the
following Section we take up the case of $\rho(g)$  at
the operator  level.

We have seen that the two distinct possibilities for the
quotient $H/H_0$ are given by eqn.(2.6).  For simplicity
in Case A we limit ourselves to $H=H_0$, ie., we will
hereafter consider just two possibilities:
     \begin{eqnarray}
     {\mbox{\bf Case (a):}}\;\;\;\;&&H = H_0\, ;\nonumber\\
     {\mbox{\bf Case (b):}}\;\;\;\;  && H/H_0 = U(1) \,.
     \end{eqnarray}

\noindent
In Case (b) we have $H\simeq H_0 \times U(1)$ apart possibly for
some global identification rules.  The intermediate case of 
$H/H_0$ discrete nontrivial can be handled by straightforward
modifications of the analysis to follow.  In Case (a)
the coset spaces $\Sigma_0, \Sigma$ coincide; and the
harmonic analysis to be now developed for functions on
$\Sigma$ to study $\psi(g)$ can later be used to study
$\rho(g)$.  In Case (b), since $H$ is `larger' than 
$H_0$ by exactly one $U(1)$ angle, the coset space 
$\Sigma_0$ is also `larger' than $\Sigma$ by (locally)
one angle variable in the range $(0,2\pi)$.  Whereas for 
$\psi(g)$ we can use the results of harmonic analysis
arising from appropriate UR's of $G$ on $\Sigma_0$ or on
$\Sigma$, for $\rho(g)$ we have to use the results on 
$\Sigma$ alone.  At this point, focussing on $\psi(g)$
we divide the discussion into Cases (a) and (b).

\subsection{Harmonic analysis in Case (a): $\;H=H_0$}

With respect to $H\subset G$ the significant information
available about the properties of the generalised coherent
state vectors $\psi(g)\,\in\,\vartheta(\psi_0)$ can be
summarised as follows;
     \begin{eqnarray}
      h\,\in\,H:\;\; {\cal D}^{(J_{0})}(h)\psi_0 &=&
      \psi_0\, ;\nonumber\\
      \psi(g) &=& {\cal D}^{(J_{0})}(g) \psi_0\, ,\nonumber\\
      \psi(gh) &=& \psi(g)\, ;\nonumber\\
      {\cal D}^{(J_{0})}(g^{\prime})\psi(g) &=&
      \psi(g^{\prime}g)\,.
      \end{eqnarray}

\noindent
Let us denote a general point on $\Sigma$, a general
$H$-coset, by $q=gH$.  The identity coset $eH=H$ is the
distinguished origin $q_0\,\in\,\Sigma$.  A general
$g^{\prime}\,\in\,G$ maps $q$ to $q^{\prime}=
g^{\prime} q$.  Also denote by $\ell(q)\,\in\,G$ a
(local) choice of coset representatives $\Sigma\rightarrow
G$ :
     \begin{eqnarray}
     q\,\in\,\Sigma \longrightarrow
     \ell(q)\,\in\,G:\;\; \ell(q) q_0 = q\,.
      \end{eqnarray}

\noindent
(In general, considering that $G$ is a principal fibre
bundle over $\Sigma$ as base and $H$ as fibre and 
structure group, such coset representatives are
definable only locally, and not in a globally smooth
way; however these aspects involving domains of definition
and overlap transition functions can be taken care of
suitably).  Then the `independent information' 
contained in the vectors $\psi(g)$ can be reexpressed
as follows:
     \begin{eqnarray}
     \psi_0(q) &=&\psi(\ell(q))\,,\;\;\;
     \psi_0(q_0)=\psi_0 \,;\nonumber\\
     {\cal D}^{(J_{0})}(g) \psi_0(q) &=&
     \psi(g\;\ell(q))\nonumber\\
     &=& \psi(\ell(gq))\nonumber\\
     &=& \psi_0(gq)\,.
      \end{eqnarray}

\noindent
Based on these relationships we set up a UR of $G$ on
functions on $\Sigma$ in this manner.  The Hilbert
space of the UR is
     \begin{eqnarray}
     L^2(\Sigma, C) = \left\{f(q)\,\in \,C\;\bigg|\;
     \int\limits_{\Sigma} d\mu(q)| f(q)|^2
     < \infty \right\}\,.
     \end{eqnarray}

\noindent
Here $d\mu(q)$ is the $G$-invariant integration volume element 
on $\Sigma, d\mu(gq) = d\mu(q)$; in the case of compact $G$ and
$H$ we assume it is normalised to unit total volume for $\Sigma$.
On these (scalar valued) functions $f(q)$ we define the action
of $G$ by unitary operators ${\cal U}(g)$ :
     \begin{eqnarray}
     ({\cal U}(g)f) (q) = f(g^{-1} q)\,.
     \end{eqnarray}

\noindent
It is now recognized that we have here the UR 
${\cal D}^{\mbox{(ind},0)}$ of $G$ induced from the 
identity or trivial one-dimensional UIR of $H$, as
described in Appendix B, eqn.(B.4).  (The superscript $0$
is a reminder that the induction is from the trivial
representation of $H$).  As explained there, by well-known
reciprocity theorems this UR ${\cal D}^{\mbox{(ind,0)}}$
of $G$ contains a general UIR ${\cal D}^{(J)}$ of $G$ 
as many times as the latter contains the trivial one
dimensional UIR of $H$.  To make this quite explicit, at
this point we choose the `magnetic quantum number' $M$
within UIR's of $G$ to consist of a triple $M=\mu\;j\;m$:
here $\mu$ is a multiplicity label for UIR's of $H$, $j$
is a label for UIR's of $H$, and $m$ is a `magnetic quantum
number' within the $j$th UIR of $H$.  (As with $J$ and $M$,
here too $j$ and $m$ in general stand for sets of several
quantum numbers each).  Then the general matrix element
within the $J$th UIR of $G$ appears, adapted to $H$, as:
     \begin{eqnarray}
     {\cal D}^{(J)}_{MM^{\prime}}(g) =
      {\cal D}^{(J)}_{\mu j m, \mu^{\prime} j^{\prime}
      m^{\prime}}(g)\,.
      \end{eqnarray}

\noindent
With this information we have the result that a complete 
orthonormal basis for the Hilbert space $L^2(\Sigma, C)$ 
is given by
     \begin{eqnarray}
      Y^{(J\lambda)}_{\mu j m} (q) &=&
      N^{1/2}_J  {\cal D}^{(J)}_{\mu j m, \lambda 00}
      (\ell(q))\; ,\nonumber\\
      Y^{(J\lambda)}_{\mu j m} (q_0) &=&
      N^{1/2}_J \delta_{\lambda \mu} \delta_{j0} \delta_{m0}\;.
      \end{eqnarray}

\noindent
(Here again $j=m=0$ corresponds to the identity UIR of $H$).
We can say that there are as many independent `spherical
harmonics' on $\Sigma$ of representation type $J$ as
${\cal D}^{(J)}$ contains $H$-scalar states, and $\lambda$
counts this multiplicity.  The basic properties of 
these functions are:
     \begin{eqnarray*}
     Y^{(J\lambda)}_{\mu j m} (gq) =
     \sum\limits_{\mu^{\prime}j^{\prime}m^{\prime}}
     {\cal D}^{(J)}_{\mu j m, \mu^{\prime}j^{\prime}
      m^{\prime}}(g)
     Y^{(J\lambda)}_{\mu^{\prime}j^{\prime}m^{\prime}}(q) \;;
     \end{eqnarray*}
     \begin{eqnarray}
     \int\limits_{\Sigma} d\mu(q) Y^{(J^{\prime}\lambda^{\prime})}
     _{\mu^{\prime}j^{\prime}m^{\prime}}(q)^*
     Y^{(J\lambda)}_{\mu j m}(q) &=&
     \delta_{J^{\prime}J} \delta_{\lambda^{\prime}\lambda}
     \delta_{\mu^{\prime}\mu} \delta_{j^{\prime}j}
     \delta_{m^{\prime}m}\; ;\nonumber\\
     \sum\limits_{J \lambda \mu j m}  Y^{(J\lambda)}_{\mu j m} (q)
     Y^{(J\lambda)}_{\mu j m} (q^{\prime})^* &=&
     \delta(q^{\prime} , q) \,.
     \end{eqnarray}

\noindent
In the last completeness relation we have the Dirac delta 
function on $\Sigma$ with respect to the volume element 
$d\mu(q)$.

Now we use the above tools to perform the harmonic
analysis of $\psi_0(q)$.  The results, as may be
expected, will be simple, but the pattern for the
later treatment of $\rho(g)$ will be set.  Let us
denote an orthonormal basis for ${\cal H}^{(J_{0})}$,
adapted to $H$, by $\Psi^{(J_{0})}_{\mu j m}$:
     \begin{eqnarray}
     {\cal D}^{(J_{0})}(g) \Psi^{(J_{0})}_{\mu j m} &=&
     \sum\limits_{\mu^{\prime}j^{\prime}m^{\prime}}
     {\cal D}^{(J_{0})}_{\mu^{\prime}j^{\prime}m^{\prime},
     \mu j m}(g)
     \Psi^{(J_{0})}_{\mu^{\prime}j^{\prime}m^{\prime}}\; ,
     \nonumber\\
     \Psi^{(J_{0})\dag}_{\mu^{\prime}j^{\prime}m^{\prime}} 
     \Psi^{(J_{0})}_{\mu j m} &=& \delta_{\mu^{\prime}\mu}
     \delta_{j^{\prime}j} \delta_{m^{\prime}m} \;.
     \end{eqnarray}

\noindent
Since $\psi_0$ is an $H$-invariant vector in ${\cal H}^{(J_{0})}$,
it follows that the UIR ${\cal D}^{(J_{0})}$ of $G$ contains at
least one $H$-scalar state.  Let us for simplicity choose
$\psi_0$ to be the one corresponding to the multiplicity label
$\mu$ having the value unity:
     \begin{eqnarray}
     \psi_0 = \Psi^{(J_{0})}_{100} \,.
     \end{eqnarray}

\noindent
Then the generalised coherent states $\psi(g)$, and hence
$\psi_0(q)$, can be written out in explicit  detail:
     \begin{eqnarray}
     \psi(g) = {\cal D}^{(J_{0})}(g)\psi_0 &=&
     \sum\limits_{\mu j m} 
     {\cal D}^{(J_{0})}_{\mu j m, 100}(g)\Psi^{(J_{0})}_{\mu j m} 
     \;;\nonumber\\
     \psi_0(q) = \psi(\ell(q)) &=&
     N^{-1/2}_{J_{0}} \sum\limits_{\mu j m}
     Y^{(J_{0},1)}_{\mu j m} (q) \Psi^{(J_{0})}_{\mu j m}\;.
     \end{eqnarray}

\noindent
We see that the Fourier coefficients of $\psi_0(q)$ are
very simple:
     \begin{eqnarray}
     \int\limits_{\Sigma} d\mu(q) Y^{(J\lambda)}_{\mu j m}(q)^*
     \psi_0(q) = N^{-1/2}_{J_{0}} \delta_{JJ_{0}}
    \delta_{\lambda,1} \Psi^{(J_{0})}_{\mu j m}\;.
     \end{eqnarray}

\noindent
This is as expected, and the expansion of $\psi_0(q)$ in the 
complete set $\left\{Y^{(J\lambda)}_{\mu j m}(q) \right\}$
gives back the second of eqn.(2.18).

\subsection{Harmonic analysis in Case (b):  $\;H\simeq H_0 \times U(1)$}

Now $H_0$ and $H$ are distinct.  The results expressed in 
eqns.(2.18,19) remain valid and adequate as far as the 
harmonic analysis of $\psi(g)$ or $\psi_0(q)$ is concerned;
we must just imagine $H$ and $\Sigma$ replaced throughout by 
$H_0$ and $\Sigma_0$ in the Case (a) analysis.  However
since the larger subgroup $H$ is now available, we outline
the kind of induced UR of $G$  we would have to set up
on functions on the smaller coset space $\Sigma=G/H$,
suitable for the harmonic analysis of $\psi(g)$ if one
so wished.

With respect to $H\simeq H_0 \times U(1)\subset G$, in contrast 
to the previous eqn.(2.8), we can now say the following about
the family of generalised coherent states:
     \begin{eqnarray}
     h\,\in\,H:\;\;\; {\cal D}^{(_{0})}(h) \psi_0 &=&
     e^{i\varphi(h)}\psi_0 \;,\nonumber\\
     \varphi(h^{-1})&=& -\varphi(h)\;,\nonumber\\
     \varphi(h) &=& 0 \;\mbox{for}\;
     h\,\in\,H_0 \;;\nonumber\\
     \psi(gh) &=& e^{i\varphi(h)}\psi(g) \;;\nonumber\\
     {\cal D}^{(J_{0})} (g^{\prime}) \psi(g) &=&
     \psi(g^{\prime} g) \;.
     \end{eqnarray}

\noindent
(The last statement here is the same as before).  Now let 
us denote a general $H$-coset, a point of $\Sigma$, by
$r=gH$.  (Since $H_0\neq H$, the symbol $q$ has been used 
up to label points of $\Sigma_0$).  The identity coset
$eH=H$ is the distinguished origin $r_0\,\in\,\Sigma$;
and $g^{\prime}\,\in\,G$ maps $r$ to $r^{\prime}=g^{\prime}r$.
In local coordinates, the point $q\,\in\,\Sigma_0$ (the 
larger coset space) is a pair, $q=(r,\alpha)$ where 
$r\,\in\,\Sigma$ and $\alpha\,\in\,[0,2\pi)$ is the
U(1) angle.  Now let $\ell(r)\,\in\,G$ be a choice
of (local) coset representatives $\Sigma\rightarrow G$:
     \begin{eqnarray}
     r\,\in\,\Sigma \rightarrow \ell(r)\,\in\,
     G:\;\; \ell(r) r_0 = r\,.
     \end{eqnarray}

\noindent
Then the information (2.20) about the generalised coherent 
states $\psi(g)$ gets expressed in this way:
     \begin{eqnarray}
      \tilde{\psi}_0(r) = \psi(\ell(r)) \,,\;&&\;
      \tilde{\psi}_0(r_0) =\psi_0\; ;\nonumber\\
      {\cal D}^{(J_{0})} (g) \tilde{\psi}_0(r)
      &=&  {\cal D}^{(J_{0})}(g\ell(r))\psi_0\nonumber\\
       &=&  {\cal D}^{(J_{0})} (\ell(gr) \ell(gr)^{-1}
       g\ell(r))\psi_0\nonumber\\
       &=& e^{i\varphi(\ell(gr)^{-1} g\ell(r))}
        \tilde{\psi}_0 (gr)\;.
      \end{eqnarray}

\noindent
The characteristic difference compared to eqn.(2.10), namely the
presence of the nontrivial phase factor, is to be noted.
This means that for analysing $\psi(g)$ in this setting 
we must construct a UR of $G$ on square integrable functions 
over $\Sigma$ {\em involving a nontrivial multiplier}.  
The Hilbert space of this representation is (for simplicity 
we use the same symbol $f$ as in eqn.(2..11)):
     \begin{eqnarray}
      L^2(\Sigma,C) = \left\{f(r)\,\in\,C \;\bigg|\;
      \int\limits_{\Sigma} d\nu(r)| f(r)|^2 < \infty\right\} ,
      \end{eqnarray}

\noindent
where $d\nu(r) = d\nu(gr)$ is the $G$-invariant normalised 
volume element on $\Sigma$.  (Therefore locally 
$d\mu(q)=\frac{1}{2\pi}d\nu(r) d\alpha$).  On such $f(r)$
 we set up a UR $\tilde{\cal U}(g)$ of $G$ as follows:
     \begin{eqnarray}
     (\tilde{\cal U}(g)f)(r) =
      e^{i\varphi(\ell(r)^{-1} g\;\ell(g^{-1}r))}
     f(g^{-1}r)\,.
     \end{eqnarray}

\noindent
This is recognized to be the UR of $G$ induced from the
nontrivial one dimensional UIR $e^{i\varphi(h)}$ of $H$, in which
$H_0$ is represented trivially.  One can now proceed with
the harmonic analysis of $\psi(g)$ in which the subgroup $H$
plays the key role, by starting from an orthonormal basis
for ${\cal H}^{(J_{0})}$ adapted to $H$ rather than merely
to $H_0$.  However as we have already performed the
harmonic analysis of $\psi(g)$ with respect to its strict
stability subgroup $H_0$, we do not pursue Case (b) for
$\psi(g)$ any further; these additional details will 
become relevant in the next Section, and will be spelt 
out there.
\section{Harmonic analysis for the projections}
\setcounter{equation}{0}

When we turn to an analysis of the projection operators 
$\rho(g)=\psi(g)\psi(g)^{\dag}$ we see that in both 
Cases (a) and (b) the analysis must be based on the
strict stability group $H$ of $\rho_0$, and therefore 
with the appropriate induced UR  of $G$ on functions over 
$\Sigma$.  (Thus uniformly the vector level analysis is
better done using $H_0$, and the operator level analysis
using $H$, whatever the relationship between $H_0$ and
$H$ may be).  The results of the harmonic analysis
are now not as simple as for $\psi(g)$ in eqns. (2.18,19).  
We now treat the details as far as possible parallel to 
the discussions in the previous Section, first for Case (a)
and then for Case (b).

\subsection{Projection operators in Case (a)}

The basic facts about the family of projection operators
$\rho(g)$ are, in the pattern of eqns.(2.8,20):
     \begin{eqnarray}
     h\,\in\,H:\;\;\; {\cal D}^{(J_{0})}(h)
     \rho_0 {\cal D}^{(J_{0})}(h)^{\dag} &=& \rho_0 \; ;
      \nonumber\\
      \rho(g)&=& {\cal D}^{(J_{0})}(g) \rho_0 
      {\cal D}^{(J_{0})}(g)^{\dag}\;, \nonumber\\
      \rho(gh) &=& \rho(g) \;;\nonumber\\
      {\cal D}^{(J_{0})}(g^{\prime}) \rho(g)
       {\cal D}^{(J_{0})}(g^{\prime})^{\dag} &=&
      \rho(g^{\prime}g)\;.
      \end{eqnarray}

\noindent
Using the notations for the coset space $\Sigma=G/H$ 
already introduced in the previous Section under Case(a),
and the coset representatives $\ell(q)$ in eqn.(2.9), we
can express the content of eqns.(3.1) as follows:
     \begin{eqnarray}
     \rho_0(q) = \rho(\ell(q))\,,\;\;\; \rho_0(q_0)&=&\rho_0\; ;
     \nonumber\\
     {\cal D}^{(J_{0})}(g)\rho_0(q) 
     {\cal D}^{(J_{0})}(g)^{\dag}&=& \rho(g\ell(q))\nonumber\\
     &=&\rho(\ell(gq))\nonumber\\
     &=&\rho_0(gq) \,.
     \end{eqnarray}

\noindent
For the harmonic analysis of $\rho(g)$ or $\rho_0(q)$ we 
therefore set up on $L^2(\Sigma, C)$, by eqn..(2.12), the 
induced UR ${\cal D}^{(\mbox{ind},0)}
(g)={\cal U}(g)$ of $G$ just as 
was done for $\psi(g)$ in Case (a).  The UIR contents of
this UR are as described in the previous Section.  A complete
orthonormal basis is provided by eqns.(2.14) with the 
properties (2.15); so the UIR ${\cal D}^{(J)}$ of $G$ is 
present as many times as it contains $H$-scalar states, and the 
index $\lambda$ counts this multiplicity.

We can now project out the `Fourier Coefficients'
$\rho^{J\lambda}_{\mu j m}$ of $\rho(g)$ as operators 
acting on ${\cal H}^{(J_0)}$:
     \begin{eqnarray}
     \rho^{J\lambda}_{\mu j m} = \int\limits_{\Sigma}
     d\mu(q) Y^{(J\lambda)}_{\mu j m} (q)^* 
     \rho_0(q)\,.
     \end{eqnarray}

\noindent
On the one hand combined use of eqns.(2.15,3.2) and 
unitarity of ${\cal D}^{(J)}$ leads to the expected
tensor operator behaviour:

     \begin{eqnarray}
     {\cal D}^{(J_{0})}(g) \rho^{J\lambda}_{\mu j m}
     {\cal D}^{(J_{0})}(g)^{\dag} =\sum\limits_
     {\mu^{\prime}j^{\prime}m^{\prime}}
     {\cal D}^{(J)}_{\mu^{\prime}j^{\prime}m^{\prime},
     \mu j m}(g) \rho^{J\lambda}_{\mu^{\prime}
      j^{\prime}m^{\prime}}\;.
      \end{eqnarray}

\noindent
On the other hand the completeness relation in
eqn.(2.15) gives
     \begin{eqnarray}
     \rho_0(q) =\sum\limits_{J\lambda\mu j m} Y^{(J\lambda)}
     _{\mu j m}(q) \rho^{J\lambda}_{\mu j m}\;,
     \end{eqnarray}

\noindent
while of course $\rho(g)$ for general $g$ is obtained by going to 
the $H$ coset of $g$ :
     \begin{eqnarray}
     g=\ell(q)h,\;\; q\,\in\,\Sigma\,,\;\;
     h\,\in\,H:\;\; \rho(g) =\rho_0(q) \,.
     \end{eqnarray}

\noindent
However all this by no means implies that all the operators
$\rho^{J\lambda}_{\mu j m}$ are nonvanishing.  What is clear is 
that the UIR's $J$ of $G$ that appear as tensor operators 
in the harmonic analysis of $\rho(g)$ (and their corresponding
multiplicities) must be some subset of the spectrum of UIR's
of $G$ that are known to be contained in the induced UR
${\cal D}^{(\mbox{ind},0)}\equiv{\cal U}(\cdot)$, as dictated
by the reciprocity theorem.  Indeed one can see immediately
that, when $G$ and $H$ are both compact and $G/H$ is nontrivial,
${\cal H}^{(J_{0})}$ is finite dimensional whereas
${\cal D}^{(\mbox{ind},0)}$ is infinite dimensional; therefore
only a finite subset of the $\rho^{J\lambda}_{\mu j m}$ can 
be nonzero.

To pin down further the tensor operators $\rho^{J\lambda}
_{\mu j m}$ we relate them directly to the fiducial vector
$\psi_0\,\in\,{\cal H}^{(J_{0})}$ and to the generalised
coherent states $\psi(g)$.  We have introduced in eqn.(2.16)
the orthonormal basis $\Psi^{(J_{0})}_{\mu j m}$ for ${\cal H}
^{(J_{0})}$ adapted to $H$, and in eqn.(2.17) we have identified
$\psi_0$ to be $\Psi^{(J_{0})}_{100}$.  This has given the explicit
expressions (2.18) for $\psi(g)$ and $\psi_0(q)$.  Combining these
various results and also using eqn.(2.14) we see that the integrand
on the right hand side in eqn.(3.3) is
     \begin{eqnarray}
     Y^{(J\lambda)}_{\mu j m}(q)^* \rho_0(q)&=&
     N^{1/2}_J\sum\limits_{\buildrel
     {\mu^{\prime}j^{\prime}m^{\prime}}\over
     {\mu^{\prime\prime}j^{\prime\prime}m^{\prime\prime}}}\,
     \Psi^{(J_{0})}_{\mu^{\prime}j^{\prime}m^{\prime}}
     \Psi^{(J_{0})^{\dag}}_{\mu^{\prime\prime}j^{\prime\prime}
     m^{\prime\prime}}\nonumber\\
    &&\;\;\;\times \;{\cal
D}^{(J_{0})}_{\mu^{\prime}j^{\prime}m^{\prime},100}
     (\ell(q)) \,
     {\cal D}^{(J_{0})}_{\mu^{\prime\prime}j^{\prime\prime}
     m^{\prime\prime},100} (\ell(q))^*\,
     {\cal D}^{(J)}_{\mu j m,\lambda 00}
     (\ell(q))^* \,.
     \end{eqnarray}

\noindent
For the product of the two ${\cal D}^*$ matrix elements we 
have the Clebsch-Gordan decomposition given in eqn.(A.7)
involving the Clebsch-Gordan coefficients of $G$ adapted to $H$:
     \begin{eqnarray}
     {\cal D}^{(J_{0})}_{\mu^{\prime\prime}j^{\prime\prime}
      m^{\prime\prime},100}
     (\ell(q))^*&&\!\!\!\! 
     {\cal D}^{(J)}_{\mu j m, \lambda 00} (\ell(q))^* \nonumber\\
     &=& \sum\limits_{\buildrel{\buildrel
      {J^{\prime}\Lambda}\over {\nu k n}}\over{\nu^{\prime}
      k^{\prime} n^{\prime}}}
      {\cal D}^{(J^{\prime})}_{\nu^{\prime}k^{\prime}n^{\prime},
       \nu k n}(\ell(q))^* \,
       {C^{J_{0}}_{\mu^{\prime\prime}j^{\prime\prime}
       m^{\prime\prime}} \;^{J}_{\mu j m}\; ^{J^{\prime}\Lambda}
      _{\nu^{\prime}k^{\prime}n^{\prime}}}^*\;
       C^{J_{0}}_{100}\;^{J}_{\lambda 00} \;^{J^{\prime}\Lambda}
       _{\nu k n} \nonumber\\
       &=&  \sum\limits_{\buildrel
      {J^{\prime}\Lambda \nu}\over{\nu^{\prime}
      k^{\prime} n^{\prime}}}\;
      N^{-1/2}_{J^{\prime}} \;
      {C^{J_{0}}_{\mu^{\prime\prime}j^{\prime\prime}
      m^{\prime\prime}}\; ^{J}_{\mu j m}\;
      ^{J^{\prime}\Lambda}_{\nu^{\prime}k^{\prime}n^{\prime}}}^*\;
      C^{J_{0}}_{100}\;^{J}_{\lambda 00}\;^{J^{\prime}\Lambda}
      _{\nu 00}\; Y^{(J^{\prime}\nu)}_{\nu^{\prime}k^{\prime}
      n^{\prime}} (q) \,,
      \end{eqnarray}

\noindent
since the second Clebsch-Gordan coefficient shows that in the 
sums over $k$ and $n$ only $k=n=0$ survives.  Putting (3.8)
into (3.3) and carrying out the integration we get the 
result
     \begin{eqnarray}
     \rho^{J\lambda}_{\mu j m} = \displayfrac{N_J^{1/2}}
     {N_{J_{0}}} \sum\limits_{\Lambda} \;
     C^{J_{0}}_{100}\;^{J}_{\lambda 00}\;
     ^{J_{0} \Lambda}_{100}\;
     \sum\limits_{\buildrel
     {\mu^{\prime}j^{\prime}m^{\prime}}\over
     {\mu^{\prime\prime}j^{\prime\prime}m^{\prime\prime}}}
     {C^{J_{0}}_{\mu^{\prime\prime} j^{\prime\prime}
     m^{\prime\prime}}\;^{J}_{\mu j m}\; ^{J_{0}\Lambda}_               
     {\nu^{\prime}k^{\prime}n^{\prime}}}^*
     \,\Psi^{(J_{0})}_{\mu^{\prime}j^{\prime}m^{\prime}}\;
     \Psi^{(J_{0})^{\dag}}_{\mu^{\prime\prime}j^{\prime\prime}
     m^{\prime\prime}}\;.
     \end{eqnarray}

\noindent
The sum over the outer products of the elements of the basis for
${\cal H}^{(J_{0})}$ reproduces exactly the $\Lambda$th unit tensor 
of rank  $J$ on ${\cal H}^{(J_{0})}$, as given in eqn.(A.12).
Thus we have the final result we are after:
     \begin{eqnarray}
     \rho^{J\lambda}_{\mu j m} = \displayfrac{N_{J}^{1/2}}
     {N_{J_{0}}}\; \sum\limits_{\Lambda}\; 
     C^{J_{0}}_{100}\;^{J}_{\lambda 00}\;
     ^{J_{0}\Lambda}_{100}\; 
     U^{J\Lambda}_{\mu j m}\;.
     \end{eqnarray} 

\noindent
We immediately see that a necessary condition for
$\rho^{J\lambda}_{\mu j m}$ to be nonzero is that
the UIR ${\cal D}^{(J)}$ must occur in the direct
product ${\cal D}^{J_{0}}\times {\cal D}^{(J_{0})^*}$,
which is of course reasonable.

It is also evident that a certain rectangular matrix
for each $J$, made up of specific Clebsch-Gordan  
coefficients, plays an important role here.  We may write 
(3.10) as
     \begin{eqnarray}
     \rho^{J\lambda}_{\mu j m} &=& \sum\limits_{\Lambda}
     \;\pi^{(J)}_{\lambda \Lambda}\;U^{J\Lambda}_{\mu j m}\;,
      \nonumber\\
      \pi^{(J)}_{\lambda\Lambda}&=& \displayfrac{N_{J}^{1/2}}
      {N_{J_{0}}}\;C^{J_{0}}_{100}\;
       ^{J}_{\lambda 00}\;^{J_{0}\Lambda}_{100}\;.
      \end{eqnarray}

\noindent
The row index $\lambda$ gives the multiplicity of occurrence
of $H$-scalar states within the UIR ${\cal D}^{(J)}$ of $G$,
while the column index $\Lambda$ (which has no reference to $H$)
gives the multiplicity of occurrence of ${\cal D}^{(J_{0})}$
in the decomposition of the product ${\cal D}^{(J_{0})}\times
{\cal D}^{(J)}$.  The necessary and sufficient conditions,
in Case (a), for being able to express every operator
$A$ on ${\cal H}^{(J_{0})}$ as an integral over the
projections $\rho(g)$ or $\rho_0(q)$, namely as
     \begin{eqnarray}
     A = \int\limits_{\Sigma} d\mu(q)\;a(q)\;\rho_0(q) \,,
     \end{eqnarray}

\noindent
for some $c$-number function $a(q)$ depending linearly on $A$,
are now clear.  We know in advance that the set of unit 
tensor operators $U^{J\Lambda}_{\mu j m}$, with spectrum of
$J\Lambda$ values completely and directly determined by
${\cal D}^{(J_{0})}$ with no reference to the subgroup $H$,
form a complete trace orthogonal set of operators on 
${\cal H}^{(J_{0})}$.  Given the relations (3.11) for each 
$J$ expressing the Fourier coefficients of $\rho_0(q)$ in 
terms of these unit tensors, we must be able to invert these
relations and express each $U^{J\Lambda}_{\mu j m}$ as a 
$\Lambda$-dependent linear combination over $\lambda$
of the $\rho^{J\lambda}_{\mu j m}$.  Thus the necessary
and sufficient conditions are as follows:

(i) Each UIR  ${\cal D}^{(J)}$ of $G$ contained in the
product UR ${\cal D}^{(J_{0})}\times {\cal D}^{(J_{0})^*}$
with some multiplicity must also occur in the UR ${\cal D}
^{(\mbox{ind},0)}$ of $G$ induced from the identity
UIR of $H$, with the same or higher multiplicity.

(ii)  For each such ${\cal D}^{(J)}$, the rectangular matrix
$\pi^{(J)}$ in (3.11) must have at least as many rows as it
has columns, and it must be of maximal rank, namely equal
to the number of columns.

\subsection{Projection operators in Case (b)}

The main complication now is that $\psi_0$ and $\rho_0$ have
different strict stability groups.  We therefore have to
unavoidably introduce extra quantum numbers in the state
labels to take account of the structure $H\simeq U(1)\times
H_0$.  Further in carrying out harmonic analyses over 
$\Sigma=G/H$, we must use two different sets of complete 
orthonormal spherical harmonics, one appropriate for $\psi(g)$
and another (simpler) one for $\rho(g)$.  The increase in 
index structure in ${\cal D}$-functions, $Y$-functions and
Clebsch-Gordan coefficients are all inevitable.

A general element $h\,\in\,H$ is a pair 
$h=\left(e^{i\alpha}, h_0\right)$ where $\alpha\;
\in [0,2\pi)$ and $h_0\,\in\,H_0$ (subject 
possibly to some global identification rules).  The label
$j$ for a general UIR of $H$ is also a pair
$j=(y,j_0)$ where $y\,\in\,{\cal Z}$ is the
$U(1)$ quantum number and $j_0$ labels a UIR of $H_0$
(again here $y$ and $j_0$ may be constrained in some way).
Within the UIR $j_0$ of $H_0$ we have as before an internal
magnetic quantum number $m$.  Therefore in a basis
adapted to $H$ the matrix elements in the UIR 
${\cal D}^{(J)}$ of $G$ look like
     \begin{eqnarray}
     {\cal D}^{(J)}_{MM^{\prime}} (g) =
     {\cal D}^{(J)}_{\mu y j_0 m, \mu^{\prime} y^{\prime}
     j^{\prime}_{0} m^{\prime}} (g) \,,
     \end{eqnarray}

\noindent
with the index $\mu$ counting the number of times the UIR
$j\equiv(y,j_0)$ of $H$ is present etc.  Correspondingly we
have an orthonormal basis $\Psi^{(J_{0})}_{\mu y j_0 m}$ for
${\cal H}^{(J_{0})}$ with the transformation law
     \begin{eqnarray}
     {\cal D}^{(J_0)}(g)\Psi^{(J_{0})}_{\mu y j_0 m}  =
     \sum\limits_{\mu^{\prime} y^{\prime} j^{\prime}_0
      m^{\prime}}\;
     {\cal D}^{(J_0)}_{ \mu^{\prime} y^{\prime}
     j^{\prime}_{0} m^{\prime}, \mu y j_0 m} (g) \;
     \Psi^{(J_{0})}_{\mu^{\prime} y^{\prime} 
     j^{\prime}_0 m^{\prime}}\,.
     \end{eqnarray}

\noindent
With no loss of generality we can assume that the
fiducial vector $\psi_0$, invariant under $H_0$ but
changing under the $U(1)$ part of $H$, carries the
$U(1)$ quantum number $y=1$, and is the first such
state in case of multiplicity:
     \begin{eqnarray}
     \psi_0 = \Psi^{(J_0)}_{1100} \,.
     \end{eqnarray}    

\noindent
This replaces eqn.(2.17).  For the generalised coherent state 
we have from eqn.(3.14,15), as replacement for eqn.(2.18):
     \begin{eqnarray}
     \psi(g) &=& {\cal D}^{(J_0)}(g) \Psi^{(J_0)}_{1100}
     \nonumber\\
     &=& \;\sum\limits_{\mu y j m}{\cal D}^{(J_0)}
     _{\mu y j_0 m, 1100} (g)\;\Psi^{(J_0)}_{\mu y j_0 m}\,.
     \end{eqnarray}

For points of the coset space $\Sigma$ and coset 
representatives  we use the notations $r, \ell(r)$ already 
introduced in Section 2 under Case(b).  Now as was mentioned
earlier, on $\Sigma$ we have to employ two different
complete orthonormal sets of functions, one to handle
$\psi_0(r)$ and the other to handle $\rho_0(r)$.  This is
because two different induced UR's of $G$ are involved - 
in the $\psi$ case it is the UR ${\cal D}^{(\mbox{ind},10)}$ 
induced from the nontrivial one-dimensional UIR $j=(1,0)$
of $H$ as described in eqn.(2.24); in the $\rho$ case it 
is the UR ${\cal D}^{(\mbox{ind},00)}$ induced from the 
trivial one-dimensional UIR $j=(00)$ of $H$, analogous 
to eqn.(2.12).  The two systems of  complete orthonormal 
spherical harmonics on $\Sigma$ are:
     \begin{mathletters}
     \begin{eqnarray}
     {\cal D}^{(\mbox{ind},10)}:\;\;\;
     \tilde {Y}^{(J,\lambda)}_{\mu y j_0 m} (r) &=& 
      N^{1/2}_J \; {\cal D}^{(J)}_{\mu y j_0 m, \lambda 100}
     (\ell(r))\;;  \\
     {\cal D}^{(\mbox{ind},00)}:\;\;\;
     Y^{(J,\lambda)}_{\mu y j_0 m} (r) &=& 
     N^{1/2}_J \; {\cal D}^{(J)}_{\mu y j_0 m, \lambda 000}
     (\ell(r))\;.
     \end{eqnarray}
     \end{mathletters}

\noindent
We must appreciate that the spectrum of $(J, \lambda)$ values
present in the two cases may be different, even though each
set by itself is orthonormal and complete over $\Sigma$ with
respect to the measure $d\;\nu(r)$.  The transformation 
properties under $G$ action, orthonormality and completeness
relations in each case are analogous to eqn.(2.15) and 
need not be repeated.

Equations (3.1) continue to hold, while we replace eqn.(3.2)
and the second of eqns.(2.18) by:
     \begin{eqnarray*}
     \rho_0(r) = \rho(\ell(r)) &=& \tilde{\psi}_0(r)
     \tilde{\psi}_0(r)^{\dag},\\
     \rho_0(r_0)&=& \rho_0 \,;
     \end{eqnarray*}
   
      \[{\cal D}^{(J_0)}(g) \rho_0(r) {\cal D}^{(J_0)}(g)^{\dag}
      = \rho_0(gr) \,;\]

      \begin{eqnarray}   
      \tilde{\psi}_0(r) = \psi(\ell(r)) &=&\sum\limits_{\mu y j_0 m}\;
      {\cal D}^{(J_0)}_{\mu y j_0 m, 1100}(\ell(r))\;
      \Psi^{(J_0)}_{\mu y j_0 m}\nonumber\\
      &=& N^{-1/2}_{J_{0}}\sum\limits_{\mu y j_{0} m}\;
      \tilde{Y}^{(J_{0},1)}_{\mu y j_0 m} (r)
      \Psi^{(J_0)}_{\mu y j_0 m}\,.
      \end{eqnarray}

\noindent
The pattern of calculations from here onwards is similar to Case(a).
We define the Fourier coefficients of the projection operators
$\rho_0(r)$ with respect to the basis (3.17b) as
     \begin{eqnarray}
      \rho^{J\lambda}_{\mu y j_0 m} &=& \int\limits_{\Sigma}
      d\nu(r)\;\
      Y^{(J,\lambda)}_{\mu y j_0 m} (r)^*\;
      \rho_0(r) \,,\nonumber\\
      \rho_0(r) &=& \sum\limits_{J\lambda \mu y j_0 m}
      \;Y^{(J,\lambda)}_{\mu y j_0 m} (r)\;
      \rho^{J\lambda}_{\mu y j_0 m}\,;\nonumber\\
      {\cal D}^{(J_0)} (g)\; \rho^{J\lambda}_{\mu y j_0 m}\;
      {\cal D}^{(J_0)}(g)^{\dag} &=& 
      \sum\limits_{\mu^{\prime}y^{\prime}j^{\prime}_0
      m^{\prime}}\;
      {\cal D}^{(J)}_{\mu^{\prime}y^{\prime}j^{\prime}_0
      m^{\prime}, \mu y j_0 m}(g)\;
      \rho^{J\lambda}_{\mu^{\prime}y^{\prime}j^{\prime}_0
      m^{\prime}}\,.
      \end{eqnarray}

\noindent
We then use eqn.(3.18) to directly relate $\rho^{J\lambda}
_{\mu y j_0 m}$ to outer products of the basis vectors of
${\cal H}^{(J_{0})}$, and then to the complete set of
unit tensors on ${\cal H}^{(J_{0})}$.  Skipping the 
intermediate steps, the final result replacing
eqn.(3.11) in Case (a) is:
     \begin{eqnarray*}
     \rho^{J\lambda}_{\mu y j_0 m} = \sum\limits_{\Lambda}\;
     \pi^{(J)}_{\lambda\Lambda}\; U^{J\Lambda}_{\mu y j_0 m} \,,
     \end{eqnarray*}

     \begin{eqnarray*}
     \pi^{(J)}_{\lambda\Lambda} = \displayfrac{N_{J}^{1/2}}
     {N_{J_{0}}}\;C^{J_{0}}_{1100}\;^{J}_{\lambda 000}\;
      ^{J_{0}\Lambda}_{1100} \,,
     \end{eqnarray*}
 
     \begin{eqnarray}
     U^{J\Lambda}_{\mu y j_{0}  m} =
     \sum\limits_{\buildrel{\mu^{\prime}y^{\prime}
     j^{\prime}_{0} m^{\prime}}\over{\mu^{\prime\prime}
     y^{\prime\prime}  j^{\prime\prime}_{0}  m^{\prime\prime}}}\;
     {C^{J_{0}}_{\mu^{\prime\prime}
     y^{\prime\prime}j^{\prime\prime}_{0}m^{\prime\prime}}\;
      ^{J}_{\mu y j_{0} m}\;
      ^{J_{0}\Lambda}_{\mu^{\prime}y^{\prime}j^{\prime}_{0} 
       m^{\prime}}}^*\;
     \Psi^{(J_{0})}_{\mu^{\prime}y^{\prime}
     j^{\prime}_{0} m^{\prime}}\;
     \Psi^{(J_{0})^{\dag}}_{\mu^{\prime\prime}
     y^{\prime\prime}j^{\prime\prime}_{0}m^{\prime\prime}}\,.
     \end{eqnarray}
   
\noindent
(For simplicity we have used the same symbols $\pi, U$ here as 
in Case (a)).  The necessary and sufficient conditions
to be able to express any operator $A$ on ${\cal H}^{(J_{0})}$
as an integral over the projections $\rho(g)=\psi(g)\psi(g)^{\dag}$
are now seen to read the same as in Case (a), except that the 
family of rectangular matrices $\pi^{(J)}$ is specified in a
different manner, and in condition (i) we have to read
${\cal D}^{(\mbox{ind},00)}$ in place of ${\cal D}^{(\mbox{ind},0)}$.
For complete clarity, we state the two conditions explicitly: 
(i) Each UIR ${\cal D}^{(J)}$ of $G$ contained in the product
UR ${\cal D}^{(J_{0})} \times {\cal D}^{(J_{0})^*}$ of $G$
with some multiplicity must also occur in the UR 
${\cal D}^{(\mbox{ind},00)}$ of $G$ induced from the identity
UIR of $H\simeq U(1)\times H_0$, with the same or
higher multiplicity.  (ii) For each such ${\cal D}^{(J)}$, the 
rectangular matrix $\pi^{(J)}$ in (3.20) must have at least
as many rows as it has columns, and it must be of maximal
rank, namely equal to the number of columns.

In concluding this Section we point out that we have made
convenient choices of the vector $\psi_0$ in terms of a
basis in ${\cal H}^{(J_{0})}$, and this must be kept in mind 
since expressions for standard Clebsch-Gordan coefficients 
where available may differ from the ones needed in (3.11,20).

\section{Applications to $SU(2)$ and $SU(3)$}
\setcounter{equation}{0}

As examples of the criteria developed in the last Section 
for the existence of the diagonal coherent state
representation for operators (in short, diagonal 
representation), we consider here some illustrative 
instances involving the simplest compact groups
$SU(2)$ and $SU(3)$.  Since the representation theory
of these groups, their Clebsch-Gordan series and (at 
least for $SU(2))$ the Clebsch-Gordan coefficients are all 
well known, we describe very briefly the main features
of each case considered.  One point worth repeating is
that the Clebsch-Gordan coefficients which appear in the
criteria for existence of the diagonal representation 
through the matrices $\pi^{(J)}$ are generally noncanonical.
We must bear in mind the use of bases for UIR's of $G$
adapted to the subgroup $H$ determined by $\psi_0$, and the 
identifications of $\psi_0$ in eqns.(3.15, 2.17).  We look 
at three $SU(2)$ cases and two $SU(3)$ cases to illustrate
the ideas.

\subsection{$SU(2)$~ Examples} 

 With $G=SU(2)$, the Clebsch-
Gordan series multiplicity label $\Lambda$ is absent, so we 
can set $\Lambda=1$ everywhere.  The UIR label $J$ has values
$0, 1/2, 1,\cdots$  with ${\cal H}^{(J)}$ being of dimension
$N_J=(2J+1)$.  We denote the generators by $T_1, T_2, T_3$.
In discussing stability subgroups we pay attention only to
the components continuously connected to the identity.\\

\noindent
{\em Example 1}:  Assume $J_0\geq 1$, and take $\psi_0$ 
to be a generic vector in ${\cal H}^{(J_{0})}$, not an
eigenvector of $\hat{n}\cdot T$ for any $\hat{n}\,\in\,S^2$.
Independently of $\psi_0$, the spectrum of unit tensor operators 
on ${\cal H}^{(J_{0})}$ is $J=0,1,2,\ldots,2 J_0$, once each.
The stability groups are $H_0=H=\{e\}$, so we have Case(a).
The orbits $\vartheta(\psi_0), \vartheta(\rho_0)$ and the
two coset spaces $\Sigma_0, \Sigma$ all coincide with
$SU(2)$ (or may be $S0(3))$ and are all three dimensional.
Since $H$ is trivial, it has only the trivial one dimensional
UIR, so the induced representation ${\cal D}^{(\mbox{ind},0)}$
of $SU(2)$ is the regular representation ${\cal D}^{\mbox{(reg)}}$.
The spectrum and multiplicity of UIR's present here is $J=0,1/2,
1,\ldots,\infty,\;{\cal D}^{(J)}$ occurring $(2J+1)$ times.
Therefore condition (i) for Case (a) is obeyed.  Turning to 
condition (ii), any basis $\Psi^{(J)}_{\mu},\mu=1,2,
\ldots,2J+1$, in ${\cal H}^{(J)}$ is an $H$-adapted basis and 
$\mu$ is a multiplicity label.   We take $\psi_0=\Psi^{(J_{0})}_1$
in ${\cal H}^{(J)}$, assuming for definiteness that in each
${\cal H}^{(J)}$ we have a noncanonical basis (not 
eigenvectors of $T_3$).  The matrices $\pi^{(J)}_{\lambda\Lambda}$
of eqn.(3.11) are column vectors with $(2J+1)$ rows:
     \begin{eqnarray}
     \pi^{(J)}_{\lambda 1} =\displayfrac{\sqrt{2J+1}}{2J_{0}+1}
      \;C^{J_{0}}_{1}\;^{J}_{\lambda}\;^{J_{0}}_{1}\,,\;\;\;
      \lambda=1,2,\ldots, 2J+1\,.
      \end{eqnarray}

\noindent
(We emphasize these are not the usual Clebsch-Gordan Coefficients). 
For each $J=0, 1,\ldots,2J_0$ in the generic case we can expect 
this to be nonzero at least for one value of $\lambda$, as no 
particular symmetries or selection rules are operative.  So
condition (ii) also holds, and the diagonal representation exists.\\

\noindent
{\em Example 2}:  Assume $J_0$ is an integer $\geq 1$, 
and take $\psi_0$ to be the eigenvector of $T_3$ with
eigenvalue $M_0=0$, ie., in the canonical basis, 
$\psi_0=\Psi^{(J_{0})}_0$.  Again the spectrum of unit tensor 
operators on ${\cal H}^{(J_{0})}$ is $J=0,1,2,\ldots,2J_0$, once 
each.  The stability groups are $H_0=H=U(1)$ generated by $T_3$, 
so we have case (a) again.  Now we use the canonical basis 
$\Psi^{(J)}_M$ in every ${\cal H}^{(J)}$, as it is adapted 
to $H$; the multiplicity labels $\lambda,\mu$ are not needed, 
and can all be set equal to unity.  The orbits $\vartheta(\psi_0),
\vartheta(\rho_0)$ and the coset spaces $\Sigma_0, \Sigma$ all
coincide with $SU(2)/U(1)  =S^2$, and are all two-dimensional.
The induced UR ${\cal D}^{(\mbox{ind},0)}$ of $SU(2)$ is the 
helicity zero UR acting on functions on $S^2$, and this contains 
the UIR's $J=0,1,2,\ldots,\infty$, once each; thus condition (i)
is obeyed.  Turning to condition (ii), for each $J=0,1,\ldots,
2J_0$ we have a single number $\pi^{(J)}_{11}$ to examine,
and it is the canonical Clebsch-Gordan coefficient
     \begin{eqnarray}
     \pi^{(J)}_{11} = \displayfrac{\sqrt{2J+1}}{2J_{0}+1}\;
     C^{J_{0}}_{0}\;^{J}_{0}\;^{J_{0}}_{0}\,.
     \end{eqnarray}

\noindent
But it is known that this vanishes for $J=1,3,\ldots, 2J_{0}-1$, 
hence condition (ii) is not obeyed, and the diagonal representation 
does not exist. This interesting situation was indeed noted by 
Klauder and Skagerstam a long time ago\cite{klauderreprints}, for the case
$J_0 = 1$.\\ 

\noindent
{\em Example 3}:  Take any $J_0\geq 1/2$, and $\psi_0$ to be
an eigenvector of $T_3$ in ${\cal H}^{(J_{0})}$ with eigenvalue
$M_0\neq 0$.  Thus in the canonical basis we have $\psi_0=
\Psi^{(J_{0})}_{M_{0}}, \,|M_0| >0$.  The spectrum of unit 
tensors on ${\cal H}^{(J_{0})}$ is $J=0,1,\ldots,2J_0$; while 
the stability subgroups are $H_0=\{e\}, H=U(1)$ generated by
$T_3$, leading to Case (b).  In each ${\cal H}^{(J)}$ 
we can use the canonical basis, and the labels $\lambda,\mu$,
are not needed.  The orbit $\vartheta(\psi_0)$ and the coset 
space $\Sigma_0$ are three dimensional, while $\vartheta(\rho_{0})$ 
and $\Sigma$ are $S^2$ as in Example 2.  The induced UR
of $SU(2)$ to be used for $\rho(g), {\cal D}^{(\mbox{ind},00)}$
is again the helicity zero UR on functions on $S^2$, with the
UIR spectrum $J=0,1,2,\ldots,\infty$, once each.  So condition(i)
of Case (b) is obeyed.  For $J=0,1,\ldots,2J_0$ we have to 
now examine the canonical Clebsch-Gordan coefficient (see
eqn.(3.20))
     \begin{eqnarray}
     \pi^{(J)}_{11} = \displayfrac{\sqrt{2J+1}}{2J_{0}+1}\;
     C^{J_{0}\;J\;J_{0}}_{M_{0}\;0\;M_{0}}\;,
     \end{eqnarray}

\noindent
and as this is nonzero if $M_0\neq 0$, condition (ii) is obeyed and 
the diagonal repesentation exists.

In these three $SU(2)$ examples, condition (i) was always
obeyed; while in Example 2 alone condition (ii) was 
violated.  Now we look at two $SU(3)$ examples, in one of
which even condition (i) fails.

\subsection{$SU(3)$~ Examples}   

With $G=SU(3)$, the
Clebsch-Gordan series multiplicity label $\Lambda$ is 
generally necessary.  The UIR's are labelled by a pair
of independent integers, $J=(p,q)$, with 
${\cal H}^{(p,q)}$, having dimension $N_{(p,q)}=\frac{1}{2}
(p+1)(q+1)(p+q+2)$.  We will throughout use the canonical 
basis within each ${\cal H}^{(p,q)}$, labelled by the 
quantum numbers $I, I_3, Y$ of  the isospin $SU(2)$
and hypercharge $U(1)$ subgroups of $SU(3)$.  We will be using 
two  subgroups, namely $U(1)\times U(1)$ and $U(2)$.  
The corresponding induced UIR's of $SU(3)$, arising from
the trivial UIR's of these subgroups, have the following 
contents as deduced from the reciprocity theorem:
     \begin{eqnarray*}
     {\cal D}^{(\mbox{ind},0)}_{U(1)\times U(1)} =
     \sum\limits^{\infty}_{\buildrel{p, q=0}\over
     {p=q\;\mbox{mod}\;3}} \;\oplus \;n_{p,q}\;
      {\cal D}^{(p,q)}\; ,
     \end{eqnarray*}

     \begin{mathletters}
     \begin{eqnarray}
      n_{p,q} &=& (p+1, q+1)_< \;;\\
      {\cal D}^{(\mbox{ind},0)}_{U(2)} &=&
      \sum\limits^{\infty}_{p=0}\;\oplus\;
      {\cal D}^{(p,p)}\,.
      \end{eqnarray}
      \end{mathletters}

\noindent
We take $J_0=(1,1)$ corresponding to the eight dimensional octet or
adjoint representation.  The spectrum of unit tensor operators
on ${\cal H}^{(1,1)}$ is known to be:
     \begin{eqnarray}
     (p,q) = (0,0),\;\; (1,1),\;\; (1,1),\;\; (3,0),\;\; (0,3),\;\; (2,2)\,.
     \end{eqnarray}

\noindent
We look at two choices of $\psi_0$.\\

\noindent
{\em Example 4}:  Take $\psi_0 = \Psi^{(1,1)}_{100}$.  Then 
$H_0=H=U(1)\times U(1)$ and we have Case (a).  In the canonical
basis $\Psi^{(p,q)}_{I\;I_3 Y}$ for UIR's of $SU(3)$, $I_3$ and
$Y$ determine a (one-dimensional) UIR of $H$, so $I$ is the
multiplicity label $\lambda, \mu, \cdots$ of the general formalism.  
From eqn.(4.4a) we see that ${\cal D}^{(\mbox{ind},0)}_
{U(1)\times U(1)}$ contains $(0,0)$ once, $(1,1)$ twice,
$(3,0)$ and  $(0,3)$ once each, and (2.2) three times.
Condition (i) is then obeyed.  Turning to condition (ii),
for each of the $(p,q)$ pairs listed in eqn.(4.5) we must
examine the matrix $\pi^{(J)}_{\lambda \Lambda}=
\pi^{(p,q)}_{Im\Lambda}$.  These involve quite simple
Clebsch-Gordan coefficients of $SU(3)$, which in turn are
Clebsch-Gordan coefficients of $SU(2)$ times so-called 
isoscalar factors.  We have the following results:
     \begin{mathletters}
     \begin{eqnarray}
     (p,q)=(0,0): \;\;\pi^{(0,0)}_{0,1} &=&
     1/8\;;\\ 
     &&\nonumber\\
     (p,q)=(1,1): \;\;\pi^{(1,1)} &=&
     \frac{1}{\sqrt{8}}\left(\begin{array}{ll}
      C^{8}_{100}\;^{8}_{000}\;^{8,1}_{100}
      &C^{8}_{100}\;^{8}_{000}\;^{8,2}_{100}\\
      C^{8}_{100}\;^{8}_{100}\;^{8,1}_{100}
      &C^{8}_{100}\;^{8}_{100}\;^{8,2}_{100}\end{array}\right)\nonumber\\
       &&\nonumber\\
      &=&\frac{1}{\sqrt{8}}\left(
      \begin{array}{cc}1/\sqrt{5}&0\\
      0&0\end{array}\right) ;\\
      &&\nonumber\\
      (p,q)=(3,0)\;\mbox{\scriptsize{or}}\;(0,3): \;\;
      \pi{^{(3,0)\;\mbox{\scriptsize{or}}\;(0,3)}_{1,1}} &=&
    \frac{\sqrt{10}}{8}\;
    C{^8_{100}}\,\,{^{10\;\mbox{\scriptsize{or}}\;10^{*}}_{100}}\,\,
      {^8_{100}}\nonumber\\
      &&\nonumber\\
      &=&         
      \frac{\sqrt{10}}{8}\frac{\sqrt{30}}{15}\;
       C^{1\;1\;1}_{0\;0\;0} = 0 \;;\\
        &&\nonumber\\
       (p,q)=(2,2): \;\;
       \pi^{(2,2)} &=&
       \left(\pi^{(2,2)}_{I,1}\right)\nonumber\\
       &&\nonumber\\
       &=&\frac{\sqrt{27}}{8}\left(\begin{array}{l}
       C^{8}_{100}\;^{27}_{000}\;^{8}_{100}\\
       C^{8}_{100}\;^{27}_{100}\;^{8}_{100}\\
       C^{8}_{100}\;^{27}_{200}\;^{8}_{100}\end{array}\right)
       =\frac{\sqrt{27}}{8}\left(\begin{array}{c}
       -\sqrt{5}/45\\0\\\frac{2\sqrt{10}}{9}\;
       C^{1\;2\;1}_{0\;0\;0}\end{array}\right)\,.
       \end{eqnarray}
       \end{mathletters}     

\noindent
We see that condition (ii) fails for $(p,q)=(1,1),(3,0),(0,3)$,
so the diagonal representation does not exist.  It is noteworthy
that in some cases we have the vanishing of the isoscalar factor, 
and in other cases of the multiplying $SU(2)$ coefficient.\\

\noindent
{\em Example  5}:  Take $\psi_0=\Psi^{(1,1)}_{000}$.
Again, as $H_0=H=U(2)$, we have Case (a).  But now when we 
examine the contents of ${\cal D}^{(\mbox{ind},0)}_{U(2)}$
in eqn.(4.4b), we see that the UIR(1,1) occurs just once,
while (3,0) and (0,3) are both absent.  This means that
even condition (i) is not satisfied, and so the diagonal
representation does not exist.

\section{The Heisenberg-Weyl Group}
\setcounter{equation}{0}

The last application of our formalism is to the noncompact 
Heisenberg-Weyl (H-W) group, denoted in this Section by G.
This will be somewhat heuristic as we shall often use Hilbert
space vectors subject to delta-function normalisation, 
induced representations whose reduction into UIR's involves continuous 
integrals, etc.  The main aim is to show the relevance of the
necessary and sufficient conditions of Section 3 for existence of
the diagonal representation in this situation which underlies
the very important case of ordinary coherent states.  While
the structure of $G$ (recalled below) is quite simple, its
UIR's and the various Clebsch-Gordan series have quite delicate
properties.  We give a brief account of all these aspects.

Topologically $G$ has the structure of ${\cal R}^3$.  Its Lie 
algebra $\underline{G}$ is spanned by three elements 
$t_j,j=1,2,3$, with the Lie bracket relations
     \begin{eqnarray}
     [t_1, t_2] = t_3\,,\;\;\; [t_1\;\mbox{or}\;t_2, t_3] = 0\,.
     \end{eqnarray}

\noindent
Finite group elements and the composition law and inverses are:
     \begin{eqnarray}
     \underline{\alpha}, \underline{\beta}\in {\cal R}^3:
    \;\;\; g(\underline{\alpha})&=&\exp(\alpha_2 t_1 -
     \alpha_1 t_2 + \alpha_3 t_3)  \,,\nonumber\\
     g(\underline{\alpha})^{-1}&=& g(-\underline{\alpha})\, ;
     \nonumber\\
     g(\underline{\alpha}) g(\underline{\beta}) &=&
     g\left(\alpha_1+\beta_1, \alpha_2 +\beta_2,
     \alpha_3 +\beta_3 +\frac{1}{2}(\alpha_1\beta_2 -
     \alpha_2\beta_1)\right) .
     \end{eqnarray}

\noindent
In a UR or UIR we will write $-iT_j, T_j$ hermitian, for $t_j$,
so the generator commutation relations and unitary operators
for finite group elements are:
     \begin{mathletters}
     \begin{eqnarray}
     [T_1, T_2] = i\,T_3\,,\;\;\;[T_1\;\mbox{or}\;T_2,\;T_3]= 0\,;\\
     g(\underline{\alpha})\longrightarrow {\cal D}
      (\underline{\alpha}) =\exp
     (i(\alpha_1 T_2 - \alpha_2 T_1 -\alpha_3 T_3)) \,.
     \end{eqnarray}
     \end{mathletters}

The adjoint action on the generators is
     \begin{eqnarray}
     {\cal D}(\underline{\alpha})\, (T_1, T_2, T_3)\,
     {\cal D}(\underline{\alpha})^{-1} =
     (T_1 + \alpha_1 T_3, T_2 + \alpha_2 T_3, T_3)\,.
      \end{eqnarray}

The UIR's of $G$ are of two types, depending on whether $T_3$ 
(which in any case is a scalar in a UIR) is zero or nonzero.  If
$T_3=0$, the UIR is one dimensional and is determined by 
choices of numerical values for $T_1, T_2$:
     \begin{eqnarray}
     {\cal D}^{(q_0,p_0)},\;\;\; (q_0,p_0)\in {\cal R}^2:\;\;\;
     T_1^{(q_0,p_0)} =q_0\,,\;\;\; T_2^{(q_0,p_0)} = p_0\,,\;\;\;
      T_3^{(q_0,p_0)} = 0\,.
     \end{eqnarray}

\noindent
On the other hand, for $T_3=c\neq 0$, by the  Stone-von Neumann
theorem we have an infinite dimensional UIR on $L^2({\cal R})$,
acting on Schr\"{o}dinger wave functions $\psi(q)$ of a real 
variable $q\in{\cal R}$ as follows:
     \begin{eqnarray}
      {\cal D}^{(c)},\;\;\; c\neq 0:&&\;\; {\cal H}^{(c)} =
      L^2({\cal R}) \,,\nonumber\\
      T_1^{(c)}&=& \hat{q} = q\,,\;\;\; T^{(c)}_2 =\hat{p} =
      -i\;c\frac{\partial}{\partial q},\;\; T_3^{(c)} = c\, .
      \end{eqnarray}

\noindent 
Thus there is an ${\cal R}^2$ - worth collection of inequivalent
one-dimensional UIR's ${\cal D}^{(q_0,p_0)},\; \dim\;{\cal H}
^{(q_0,p_0)}=1$; and an ${\cal R}-\{0\}$ worth collection of
inequivalent infinite-dimensional UIR's ${\cal D}^{(c)},\;
\dim\;{\cal H}^{(c)} = \infty$.   Every UIR is nonfaithful.

In the sequel, whenever there is no danger of confusion,
we omit the UIR labels $(q_0,p_0)$ or $c$ on the generators
$T_j$.

Turning to the Clebsch-Gordan problem, this is easily
analysed by examining the sums of the individual generators
of any two UIR's.  There are three cases to consider.
The following two results are obvious:
     \begin{mathletters}
     \begin{eqnarray}
     {\cal D}^{(q_0,p_0)} \times {\cal D}^{\left(q^{\prime}_0, 
     p^{\prime}_0\right)}&=& {\cal D}^{\left(q_0 + q_0^{\prime},
     p_0+p_0^{\prime}\right)}\;;\\
      {\cal D}^{(q_0,p_0)} \times {\cal D}^{(c)} &=&
     {\cal D}^{(c)} \;.
    \end{eqnarray}
    \end{mathletters}

\noindent
(In the latter case we may in fact appeal to eqn.(5.4)).  
In the case of ${\cal D}^{(c)}\times {\cal D}^{(c^{\prime})}$ 
we must distinguish between $c + c^{\prime} = 0$ and 
$c+ c^{\prime}\neq 0$.  In either case the generators of the 
product, acting on $L^2({\cal R}^2)$, are:
     \begin{eqnarray}
     T_1 = q + q^{\prime}\,,\;\;\; T_2 = -i\;c\frac{\partial}{\partial q}
     -i\;c^{\prime}\frac{\partial}{\partial q^{\prime}}\,,\;\;\;
     T_3 = c + c^{\prime} \,.
     \end{eqnarray}

\noindent
For $c+c^{\prime}\neq 0$ we switch to the independent 
variables $Q=q+q^{\prime}, Q^{\prime}=cq^{\prime}
-c^{\prime} q$, so 
     \begin{eqnarray}
     T_1 = Q\,,\;\;\; T_2 = -i(c+c^{\prime})\frac{\partial}
     {\partial Q}\,\,,\;\;\;
     T_3 = c+ c^{\prime} \,.
      \end{eqnarray}

\noindent
We see that $Q^{\prime}$ is totally absent and commutes with all
the $T_j$.  In case $c+c^{\prime}=0$ we have
     \begin{eqnarray}
     T_1 = q + q^{\prime}\,,\;\;\; T_2 = - i\,c\left(\frac{\partial}
     {\partial q} - \frac{\partial}{\partial q^{\prime}}
     \right) \,,\;\;\; T_3 = 0 \,,
     \end{eqnarray}

\noindent
(reminiscent of the EPR situation), and $T_1$ and $T_2$ form
a complete commuting set.  From all these results we see that
     \begin{mathletters}
     \begin{eqnarray}
      {\cal D}^{(c)}\times {\cal D}^{(-c)} &=& 
      {\int\int\atop {\cal R}^{2}}\oplus \;
      dq_0 dp_0 {\cal D}^{(q_0,p_0)}\; ;\\
      {\cal D}^{(c)}\times {\cal D}^{(c^{\prime})} &=&
      {\int\atop {\cal R}}\oplus \;
      dQ^{\prime} \cdot {\cal D}^{(c+c^{\prime})}\,,\;\;\;
      c+c^{\prime}\neq 0 .
      \end{eqnarray}
      \end{mathletters}

\noindent
In (5.11a) each one dimensional UIR ${\cal D}^{(q_0,p_0)}$
appears once in a continuous fashion; while in (5.11b) the
single infinite dimensional UIR ${\cal D}^{(c+c^{\prime})}$
appears infinitely often in a continuous sense, with $Q^{\prime}$ 
being a continuous multiplicity label.  The full set of 
results for the Clebsch-Gordan problem is thus contained
in eqns.(5.7,11).

Now let us work within a particular UIR ${\cal D}^{(c)}$ acting
on ${\cal H}^{(c)}$.  From the results of the Clebsch-Gordan 
problem we see that the spectrum of irreducible `unit tensors'
definable on ${\cal H}^{(c)}$ consists only of tensors belonging
to the one-dimensional UIR's ${\cal D}^{(q_0,p_0)}$, once each in 
a continuous sense for every $(q_0,p_0)\in {\cal R}^2$.  This is
because ${\cal D}^{(c)}\times {\cal D}^{(c^{\prime})}$ never
contains ${\cal D}^{(c)}$, and ${\cal D}^{(c)}\times{\cal D}
^{(q_0,p_0)}$ is exactly ${\cal D}^{(c)}$.  These unit tensors
are the familiar H-W displacement operators which are a subset of 
the unitary ${\cal D}^{(c)}(\underline{\alpha})$ themselves.
The displacement operators are
     \begin{eqnarray}
     D^{(c)}(\alpha_{\perp})={\cal D}^{(c)}(\alpha_{\perp},0) =
     \exp\left(i\;\alpha_1\hat{p} -i\;\alpha_2 \hat{q}\right),
     \end{eqnarray}

\noindent
and for them the (finite form of the) adjoint action is
     \begin{eqnarray}
     {\cal D}^{(c)}(\underline{\beta})
      D^{(c)}(\alpha_{\perp}) {\cal D}^{(c)}
     (\underline\beta)^{-1} = e^{ic(\alpha_1\beta_2-\alpha_2\beta_1)}
      D^{(c)}(\underline{\alpha}_{\perp})\,.
      \end{eqnarray}

\noindent
Therefore for each $(q_0,p_0)\in {\cal R}^2$ we define the 
(unitary) unit tensor operator
     \begin{eqnarray}
     U^{(q_0,p_0)} = D^{(c)} \left(\frac{q_0}{c},\frac{
      p_0}{c}\right) .
     \end{eqnarray}

\noindent
(For simplicity we avoid the label $c$ on these
operators).  Then from (5.13) we see that they have the 
correct transformation property, ie. they belong to
the one-dimensional UIR's ${\cal D}^{(q_0,p_0)}$:
     \begin{eqnarray}
     {\cal D}^{(c)}(\underline{\alpha})
     U^{(q_0,p_0)} {\cal D}^{(c)}(\underline{\alpha})^{-1}
     &=& e^{i(\alpha_1 p_0 - \alpha_2 q_0)}
      \;U^{(q_0,p_0)}\nonumber\\
     &=&{\cal D}^{(q_0,p_0)}(\underline{\alpha})\;
     U^{(q_0,p_0)} .
     \end{eqnarray}

\noindent
Moreover by familiar calculations, say in a basis of
eigenvectors of $\hat{q}$, we can verify the trace
orthonormality property in the delta function sense:
     \begin{eqnarray}
     \mbox{Tr}\left(U^{\left(q^{\prime}_0,p^{\prime}_0\right)
     ^{\dag}} U^{(q_0,p_0)}\right) =
      2\,\pi\,c\,\delta\left(q_0^{\prime}-q_0\right)
      \delta\left(p^{\prime}_0 - p_0\right) .
      \end{eqnarray}

\noindent
A general Hilbert-Schmidt operator $A$ on ${\cal H}^{(c)}$ can 
then be expanded as an integral over these unit tensors:
     \begin{eqnarray}
     A &=&  {\int\int\atop {\cal R}^{2}}
     \; dq_0 \;dp_0 \;a(q_0,p_0) \;
     U^{(q_0,p_0)}\; ,\nonumber\\
     a(q_0,p_0)&=& \frac{1}{2\pi c}\;\mbox{Tr}\;
     \left(U^{(q_0,p_0)^{\dag}}\, A\right) ,\nonumber\\
     \mbox{Tr}(A^{\dag}A) &=& 2\pi c  {\int\int\atop 
     {\cal R}^{2}}\;
     dq_0\; dp_0 |a(q_0,p_0)|^2 \,.
     \end{eqnarray}

\noindent
This is the Weyl representation for operators, and eqns.(5.15,
16,17) are the analogues in the present case for eqns.(A.13,14,
15) of the compact group case.  All these results are available
in advance of the choice of a fiducial vector, construction of
its generalised coherent states, etc.

Now choose a fiducial unit vector $\psi_0\in{\cal H}^{(c)}$.
From elementary quantum mechanics it is known that every real
linear combination of $\hat{q}$ and $\hat{p}$ has a continuous
spectrum and hence no normalisable eigenvectors.  Therefore
the stability group $H_0$ of $\psi_0$ is trivial.  On the other 
hand, $H$ is ${\cal R}$ (but effectively just $U(1)$) with
generator $T_3$:
     \begin{eqnarray}
     \psi_0\in {\cal H}^{(c)}\,,\;\; \|\psi_0\| = 1:\;\;\; H_0
       =\{e\}\,,\;\;\;
     H = \left\{e^{-i\alpha_3 c},\; \alpha_3 \in {\cal R}\right\}\,.
      \end{eqnarray}

\noindent
Thus for any  $\psi_0$ we have Case (b), and we have to
examine the UIR content of the UR ${\cal D}^{(\mbox{ind,00})}$ of 
$G$ induced from the trivial one dimensional UIR of $H$
(namely, $T_3=0$).  To apply the reciprocity theorem, we have
to ask how often each UIR of $G$ contains the trivial UIR of $H$.
Clearly each ${\cal D}^{(q_0,p_0)}$ contains it once, while
each ${\cal D}^{(c)}$ does not contain it at all.  In other words,
     \begin{eqnarray}
     {\cal D}^{(\mbox{ind,00})} = {\int\int\atop {\cal R}^{2}}
      \oplus \;dq_0\;dp_0\;{\cal D}^{(q_0,p_0)}\; ,
     \end{eqnarray}

\noindent
which matches exactly with the spectrum and multiplicity 
of irreducible tensor operators $U^{(q_0,p_0)}$ definable
on ${\cal H}^{(c)}$, so Condition (i) is satisfied.  As
for Condition (ii), since ${\cal D}^{(c)}\times {\cal D}
^{(q_0,p_0)} ={\cal D}^{(c)}$, the quantity $\pi^{(q_0,p_0)}$
of eqn.(3.20) is just one number (disregarding the
$N_{J_{0}}$ in the denominator), and the question is whether 
it is always nonvanishing - we examine this more directly as 
follows.

The generalised coherent states and projection operators 
arising from $\psi_0$ are:
     \begin{eqnarray}
     \psi(\underline{\alpha}) &=& {\cal D}^{(c)}(\underline
     {\alpha}) \psi_0 = e^{-ic\alpha_3}
     D^{(c)}(\alpha_{\perp}) \psi_0\; ,\nonumber\\
     \rho(\alpha_{\perp})&=& \psi(\underline{\alpha})
     \psi(\underline{\alpha})^{\dag} =
     D^{(c)}(\alpha_{\perp}) \rho_0 
     D^{(c)}(\alpha_{\perp})^{\dag} \,,\nonumber\\
     \rho_0&=& \psi_0\;\psi_0^{\dag}\; .
     \end{eqnarray}

\noindent
Under adjoint action we have
      \begin{eqnarray}
     {\cal D}^{(c)}(\underline{\beta})\,\rho(\alpha_{\perp})\,
     {\cal D}^{(c)}(\underline{\beta})^{-1} =
     \rho(\alpha_{\perp} + \beta_{\perp}) \;,
     \end{eqnarray}

\noindent
and $\beta_3$ is absent on the right.  If we denote the
Fourier transform of $\rho(\alpha_{\perp})$ by
     \begin{eqnarray}
       \tilde{\rho}(q_0,p_0) =\frac{1}{2\pi}
       {\int\int\atop {\cal R}^{2}}\;
       d^2\alpha \;
      e^{-i(\alpha_1 p_0 -\alpha_2 q_0)} \rho(\alpha_{\perp})\; ,
      \end{eqnarray}

\noindent
then (5.21) becomes
     \begin{eqnarray}
     {\cal D}^{(c)}(\underline{\beta}) \;\tilde{\rho}
     (q_0,p_0) {\cal D}^{(c)}(\underline{\beta})^{-1}
      &=& e^{i(\beta_1 p_0 -\beta_2 q_0)}
       \;\tilde{\rho}(q_0,p_0)\nonumber\\
      &=&{\cal D}^{(q_0,p_0)}(\underline{\beta})
      \;\tilde{\rho}(q_0,p_0) .
      \end{eqnarray}

\noindent
Thus each $\tilde{\rho}(q_0,p_0)$ is a tensor operator 
of type ${\cal D}^{(q_0,p_0)}$, consisting of just one
component, so it must be a scalar multiple of the unit
tensor $U^{(q_0,p_0)}$.  This factor is easily computed 
by a trace calculation since by (5.17) the unit tensors 
are a complete orthonormal (in the continuous sense) set.  
An elementary calculation shows that
     \begin{eqnarray}
     \mbox{Tr} \left(U^{\left(q^{\prime}_0, p^{\prime}_0\right)
     ^{\dag}} \tilde{\rho}(q_0, p_0)\right) =
     2\pi\left(\psi_0, U^{(q_0,p_0)}\psi_0\right)^*
     \delta\left(q_0-q_0^{\prime}\right)
     \delta\left(p_0-p_0^{\prime}\right),
     \end{eqnarray}

\noindent
which gives the result
     \begin{eqnarray}
     \tilde{\rho}(q_0,p_0) =\frac{1}{c}
     \left(\psi_0, U^{(q_0,p_0)} \psi_0\right)^*
     U^{(q_0,p_0)}\;.
     \end{eqnarray}

\noindent
The necessary and sufficient condition for the existence 
of the diagonal representation in terms of the projections
$\rho(\alpha_{\perp})$ of eqn.(5.20) is now clear: the
fiducial vector $\psi_0$ must be chosen so that for all
$(q_0,p_0)\in {\cal R}^2$,
     \begin{eqnarray}
     \left(\psi_0, U^{(q_0,p_0)} \psi_0\right)&=&
     \left(\psi_0, D^{(c)}\left(\frac{q_0}{c}, \frac{p_0}{c}
     \right) \psi_0\right) \neq 0\; ,\nonumber\\
     \mbox{ie},\;\;\; \left(\psi_0,D^{(c)}(q_0,p_0)\psi_0\right)
     &=& \left(\psi_0, e^{i\left(q_0\hat{p}-p_0\hat{q}\right)}
     \psi_0\right)  \neq 0 \;.
     \end{eqnarray}

\noindent
Assuming this condition is satisfied, we can start from the
Weyl representation (5.17) for any (Hilbert-Schmidt) operator 
$A$ and obtain from it a diagonal coherent state representation:
     \begin{eqnarray}
     A&=&  {\int\int\atop {\cal R}^{2}}\;
     dq_0\;dp_0\;
     a(q_0,p_0)\; U^{(q_0,p_0)}\nonumber\\
     &=& {\int\int\atop {\cal R}^{2}} \;
     dq_0\;dp_0\;a(q_0,p_0)\;c
     \left(\psi_0, U^{(q_0,p_0)^{\dag}}\psi_0\right)^{-1}
     \;\tilde{\rho}(q_0,p_0)\nonumber\\
     &=& {\int\int\atop {\cal R}^{2}}\;
      d^2 \alpha\;
     \phi(\alpha_{\perp}) \rho(\alpha_{\perp})\; ,\nonumber\\
     \phi(\alpha_{\perp})&=&
     \frac{c}{2\pi} {\int\int\atop {\cal R}^{2}}\;
     dq_0\;dp_0\;e^{i(\alpha_2 q_0 - \alpha_1 p_0)}
     \;a(q_0,p_0)/\left(\psi_0, U^{(q_0,p_0)}\psi_0\right)^* \,  .
     \end{eqnarray}

\noindent
From eqn.(5.17) we know that for Hilbert-Schmidt $A$, the function 
$a(q_0,p_0)$ is square integrable over ${\cal R}^2$; in relation
to this, the nature of the weight function $\phi(\alpha_{\perp})$
in the diagonal representation is determined by the factor
$\left(\psi_0,U^{(q_0,p_0)}\psi_0\right)^*$ in the denominator.

As an application we consider the case of the usual coherent 
states obtained when the fiducial vector $\psi_0$ is the Fock
vacuum or the harmonic oscillator ground state.  (Further,
for simplicity we now set $c=1$).  The wave function is
     \begin{eqnarray}
     \psi_0(q) = \pi^{-1/4}\;e^{-q^2/2} \;,
     \end{eqnarray}

\noindent
and a simple calculation gives the displacement operator
expectation value needed in eqns.(5.26,27):
     \begin{eqnarray}
     \left(\psi_0, U^{(q_0,p_0)} \psi_0\right) &=&
     \left(\psi_0, e^{i\left(q_0\hat{p} -p_0\hat{q}\right)}
     \psi_0\right)\nonumber\\
     &=& e^{-\frac{1}{4}\left(q^2_0+p^2_0\right)}\;.
     \end{eqnarray}

\noindent
This is indeed everywhere nonzero over ${\cal R}^2$, so 
the condition  (5.26) for existence of the diagonal 
representation is, as expected, obeyed.  The decaying
exponential factor here means that the tensor operators
$\tilde{\rho}^{(q_0,p_0)}$  provided by the projection 
operators $\rho(\alpha_{\perp})$ differ from the
normalised unit tensors $U^{(q_0,p_0)}$ by similarly 
decaying factors:
     \begin{eqnarray}
     \tilde{\rho}(q_0,p_0) = e^{-\frac{1}{4}
     \left(q^2_0+p_0^2\right)} U^{(q_0,p_0)}\;.
     \end{eqnarray}

\noindent
It is to compensate for this diminishing norm of
$\tilde{\rho}(q_0,p_0)$ as one goes towards infinity 
in the $(q_0,p_0)$ phase plane that one finds that the
weight function $\phi(\alpha_{\perp})$, eqn.(5.27), 
has in general the character of a very singular 
distribution: the Fourier transform of 
$\phi(\alpha_{\perp})$ is (essentially) the square 
integrable amplitude $a(q_0,p_0)$ times the exploding  
Gaussian $e^{\frac{1}{4}\left(q^2_0 +q^2_0\right)}$.

Another interesting choice of fiducial state for diagonal representation
has been considered by Haake and Wilkens\cite{haake}, namely the squeezed
vacuum. The family of generalized coherent states in this case consists of 
Gaussian pure states squeezed by a fixed amount in a fixed direction in
phase space, the centre $(q_0, p_0)$ of the Gaussian being allowed to be
located at an
arbitrary point in phase space. It is easy to see that in this case
$\langle \psi_0 |D(q_0,p_0)|\psi_0 \rangle$ is nonvanishing, and the
diagonal representation once again exists:

\begin{eqnarray}
|\psi_0\rangle = S(\eta)|0\rangle,\;&&\; S(\eta) = \exp\left(
\frac{\eta}{2}\hat a ^{\dagger \,2}
 - \frac{\eta ^*}{2}\hat a ^2\right)\,; \nonumber \\
\langle \psi_0 |D(q_0,p_0)|\psi_0 \rangle
&=&\langle 0 |S(\eta)^{-1}D(q_0,p_0)S(\eta)|0 \rangle \nonumber \\
&=&\langle 0 |D(e^{\eta} q_0, e^{-\eta}p_0)|0 \rangle \nonumber \\
&=& \exp \left( -\frac{1}{4} (e^{2\eta}q_0 ^{\,2} + e^{-2\eta}p_0
^{\,2})\right)\,.
\end{eqnarray}

 Returning to the general
result (5.27) whenever $\psi_0$ is an acceptable fiducial vector, we can
appeal to the fact that the Stone-von Neumann UIR
of the H-W group is square integrable and 
conclude that $\left(\psi_0,U^{(q_0,p_0)}\psi_0\right)$
is a square integrable function of $(q_0,p_0)$.  Thus
this amplitude must approach zero as we move far away
from the origin in ${\cal R}^2$.  This has the 
consequence that, whatever the choice of $\psi_0$
(provided (5.26) holds), the weight function 
$\phi(\alpha_{\perp})$ is in general a distribution, 
since in its Fourier representation (5.27) the square
integrable amplitude $a(q_0,p_0)$ is \underline{divided}
by another square integrable amplitude.

We now make a series of statements which help in
conveying the content of the condition (5.26) and 
in forming some (admittedly incomplete) idea of the 
set of fiducial vectors $\psi_0$ whose generalised
coherent states are rich enough to allow for the
diagonal representation:

i)  If $\psi_0(q)$ is any Gaussian wavefunction, then
$\left(\psi_0, D^{(c)} (q_0,p_0)\psi_0\right)$ is
clearly a complex gaussian in $(q_0,p_0)$, so condition 
(5.26) is satisfied.

(ii)  If $\psi_0$ does/does not obey condition (5.26),
then the transform of $\psi_0$ by the unitary operator
representing any element of the metaplectic group
$Mp(2)$ also does/does not obey condition (5.26). 
This is because under conjugation by such a unitary 
operator, $D^{(c)}(q_0,p_0)$ just becomes 
$D^{(c)}\left(q^{\prime}_0,p^{\prime}_0\right)$ for
$\left(q^{\prime}_0,p^{\prime}_0\right)$ some linear
combinations of $(q_0,p_0)$.

(iii)   If either $\psi_0(q)$ or its Fourier transform
$\tilde{\psi}_0(p)$ is a function of compact support,
then condition (5.26) is definitely not obeyed, so the
diagonal representation will not exist.  This is 
because for such $\psi_0$, the quantity $\left(\psi_0, D^{(c)}
(q_0,p_0) \psi_0\right)$ vanishes outside a finite
strip parallel to the $p_0$ or to the $q_0$ axis.  We
can also see that as Fourier transforms of functions of
compact support are entire functions of a certain class,
wave functions $\psi_0(q)$ of this class violate condition
(5.26) quite strongly - indeed their Fourier transforms 
$\tilde{\psi}_0(p)$ are of compact support.

In a purely qualitative manner we can appreciate now that
Gaussian $\psi_0(q)$ and compact-supported $\psi_0(q)$
(or $\tilde{\psi}_0(p)$) are in some ways diametrically
opposite from the point of view of condition (5.26).
To conclude this section we consider a set of fiducial 
vector choices where condition (5.26) is violated, though only
on a set  of measure zero in the $q_0-p_0$ plane.
This will then mean that in these cases for Hilbert-
Schmidt operators $A$ we do not have available the diagonal 
representation.

Consider the choice $|n \rangle$ for the fiducial vector $\psi_0$,
this being the $n^{th}$ excited state of the harmonic oscillator, 
for $n\geq 1$. The resulting generalized coherent states are the 
{\em displaced Fock states}\cite{roy}. It is
known that the matrix element (or
better expectation value) needed in condition (5.26)
is essentially a Laguerre polynomial, thus:
     \begin{eqnarray}
     |\psi_0\rangle &=& |n\rangle :\nonumber\\
     \langle\psi_0| D^{(c)} (q_0,p_0) |\psi_0
     \rangle &=& e^{-\frac{1}{4}\left(q^2_0+p^2_0\right)}
      L_n\left(\displayfrac{q^2_0+p^2_0}{2}\right) \,.
      \end{eqnarray}

\noindent
Now, as is well known, the polynomial $L_n(x)$ has exactly
$n$ distinct real zeroes in the semi infinite interval
$0<x<\infty$, hence the condition (5.26) is satisfied
except on a discrete infinite sequence of circles in the
$q_0-p_0$ plane.  
However, these singularities which are in the finite part of the
$(q_0,p_0)$ plane are not integrable. Therefore we do not have the
possibility of the diagonal representation for the above 
choices of $\psi_0$.

Recalling condition (5.26) for the existence of the
diagonal representation, and the various examples discussed above, we are
led in the Heisenberg-Weyl case to the {\em conjecture} that condition
(5.26) is obeyed if and only if the fiducial state $\psi_0$ has Gaussian
Schrodinger wave function. This will then mean that apart from the
traditional diagonal representation and the Haake-Wilkens diagonal
representation there are no other ones for the Heisenberg-Weyl group!

\section{Concluding remarks}
\setcounter{equation}{0}

We have developed necessary and coefficient conditions for a 
set of generalised coherent states, arising from a UIR of a
compact Lie group to possess the property that a diagonal 
representation in terms of projections onto these states can be
set up for any operator on the Hilbert space of the UIR.
This has required combining several structures and properties -
harmonic analysis on coset spaces, the theory of induced
representations, the associated reciprocity theorem, and the
Clebsch-Gordan problem and coefficients for the UIR's of the 
group under consideration.  Each of these plays a crucial role
in arriving at the complete set of conditions.  The explicit
examples involving $SU(2), SU(3)$ and even the Heisenberg-
Weyl group show how our conditions operate in practice, and
how we cannot do without any of the ingredients mentioned above.  
In particular it is important to appreciate that the examples 
where the diagonal representation fails to exist are not 
particularly exotic or contrived; and we can often see in
advance those cases where it is bound to be absent.

The comprehensive work of Brif and Mann\cite{Brim}  attempts also to
exploit the methods of harmonic analysis on coset spaces
to tackle the general closely related problems of Wigner
distributions and state reconstruction problems.  However,
in the absence of detailed knowledge of the irreducible 
representation contents of various induced representations
of $G$, it is easy to miss the fact that there are quite
stringent conditions to be met before a diagonal 
representation can exist.  The particular qualitative
points to be made in connection with our approach are: 
for a given UIR of $G$, the complete set of irreducible
unit tensor operators on the Hilbert space is immediately
fixed, prior to construction of any set of generalised
coherent states.  As one then considers various choices of
the fiducial vector $\psi_0$, one can see that for
$\lq{larger}\rq$ stability groups $H_0$ and $H$, the
corresponding coset spaces $\Sigma_0$ and $\Sigma$ are 
$\lq{smaller}\rq$, with the consequence that the set of
projection operators onto the generalised coherent states 
also becomes $\lq{smaller}\rq$, and so the diagonal 
representation is less likely to exist.

Finally we may mention that the issue of reproducing various
marginal probability distributions out of a Wigner like 
distribution description of density operators has played no 
role in our considerations.  This, the applolication of our methods to
phase space description of quantum systems, quantum state reconstruction
(tomography),  and other aspects
of Wigner distributions for quantum mechanics on Lie groups will be
systematically studied elsewhere.\\

\def\theequation{A.\arabic{equation}}
\appendix{\bf{Appendix A}: Notations for group representations,
Clebsch-Gordan coefficients, and unit tensors\\}
\setcounter{equation}{0}

In this Appendix we collect some items of notation and
familiar facts concerning the representation theory of 
compact groups, their Clebsch-Gordan series and coefficients
in a general case involving multiplicity, and the definition and
properties of unit tensors.  All these are used in the main
body of the paper.

We shall deal with a general compact semisimple Lie group $G$
of dimension $n$ (except that $U(1)$ factors will be allowed),
and a generic compact Lie subgroup $H$ of dimension $k < n$.
The various inequivalent UIR's of $G$ will be labelled by a
symbol $J$ which in general comprises a collection of 
independent quantum numbers.  The space of the $J$th UIR,
and its dimension, will be written as ${\cal H}^{(J)}$ and
$N_{J}$ respectively.  Within the UIR we use the label
$M$ for a complete set of state labels for an orthonormal
basis, denoting again several independent quantum numbers.
The matrix elements of the UIR matrices ${\cal D}^{(J)}$ 
are written as ${\cal D}^{(J)}_{MM^{\prime}}(g), g\,\in\,
G$.  We have:
     \begin{eqnarray}
     {\cal D}^{(J)}(g)^{\dag} \;{\cal D}^{(J)}(g) &=&
     1\;\mbox{on}\;{\cal H}^{(J)} \;,\nonumber\\
     {\cal D}^{(J)} (g_1)\;{\cal D}^{(J)}(g_2) &=&
     {\cal D}^{(J)} (g_1 g_2) \;.
     \end{eqnarray}

\noindent
The Peter-Weyl theorem gives us the orthogonality and 
completeness of these matrix elements taken from all
UIR's of $G$.  With respect to the translation invariant 
integration measure $dg$ on $G$, normalised to unit total
volume, these statements are expressed by
     \begin{mathletters}
     \begin{eqnarray}
     \int\limits_{G} dg\;{\cal D}^{(J^{\prime})}
     _{M^{\prime\prime}
     M^{\prime\prime\prime}}(g)^*{\cal D}^{(J)}
     _{MM^{\prime}}(g)&=&
     \delta_{J^{\prime}J}\delta_{M^{\prime\prime}M}
     \delta_{M^{\prime\prime\prime}M^{\prime}}/
     N_J \;,\\
     \sum\limits_{JMM^{\prime}}\;N_J\;{\cal D}^{(J)}
     _{MM^{\prime}}(g) {\cal D}^{(J)}_{MM^{\prime}}
     (g^{\prime})^* &=& \delta(g^{-1}g^{\prime})\; ,
     \end{eqnarray}
     \end{mathletters} 

\noindent
where $\delta(g)$ is the invariant Dirac delta function on $G$
with respect to $dg$.

When we consider similarly the complete family of UIR's
of the subgroup $H\subset G$, we replace the above  symbols 
with the following:

$$g \rightarrow h\,,\;\; J \rightarrow j\,,\;\; M \rightarrow m\,,\;\; 
{\cal D} ^{(J)} \rightarrow D^{(j)}\,,\;\; {\cal H}^{(J)} \rightarrow
{\cal H}^{(j)}\,,\;\mbox{and}\; N_J \rightarrow N_j \;.$$

The relations (A.2) corresponding to $H$ hold with an 
normalised integration measure $dh$, and of course
$j, m$ are again in general sets of quantum numbers.  
In particular one may ask 
for the UIR's of $G$ in a form, or in a basis, adapted to
the reduction with respect to $H$.  In that case, for each given
UIR $J$ of $G$, one has to ask which UIR's $D^{(j)}$ of
$H$ are contained within ${\cal D}^{(J)}$, and each one with what
multiplicity.  Then the state label $M$ within ${\cal D}^{(J)}$
becomes a triple $\mu j m : j$ and $m$ are the UIR and internal
state labels for $H$, while $\mu$ is an (orthonormal)
multiplicity label which distinguishes the several occurrences of
${\cal D}^{(j)}$ within ${\cal D}^{(J)}$.  If in a particular
case the multiplicity is unity, we just set $\mu =1$.  Expressed
in such a basis, the representation matrices of $G$ appear as
${\cal D}^{(J)}_{\mu j m, \mu^{\prime} j^{\prime} m^{\prime}}(g)$,
and when $g\,\in\,H$ we have
     \begin{eqnarray}
     {\cal D}^{(J)}_{\mu j m, \mu^{\prime}j^{\prime}m^{\prime}}
     (h) = \delta_{\mu^{\prime}\mu}\;
      \delta_{j^{\prime}j} \;D^{(j)}_{mm^{\prime}} (h)\,.
      \end{eqnarray}
Incidentally for the  trivial or identity representations of 
$G$ or of $H$ we write $J=0, j=0$ respectively, with $M=m=0$ as 
well.

Now we set up a notation for Clebsch-Gordan coefficients and
unit tensor operators, allowing for the possibility of 
multiplicity in the Clebsch-Gordan series, and for the 
coefficients to be complex in general.  Considering the direct 
product ${\cal D}^{(J_{1})}\times {\cal D}^{(J_{2})}$ 
of two UIR's of $G$, let the UIR 
${\cal D}^{(J_{3})}$   be present upon reduction, possibly 
several times, and introduce an orthonormal label
$\Lambda$ to distinguish its several occurrences.
Then, if $\Psi^{(J_{1})}_{M_{1}}, \Psi^{(J_{2})}_{M_{2}}$
are orthonormal bases for ${\cal H}^{(J_{1})}, 
{\cal H}^{(J_{2})}$ respectively, for each $\Lambda$ the 
product vectors
     \begin{eqnarray}
     \Psi^{(J_3,\Lambda)}_{M_3} =
     \sum\limits_{M_1,M_2}\;
     C^{J_1}_{M_1}\;^{J_2}_{M_2}\;^{J_3\Lambda}_{M_3}\;
      \Psi^{(J_1)}_{M_1}\; \Psi^{(J_2)}_{M_2}
     \end{eqnarray}

\noindent
transform by the UIR $J_3$ of  $G$, and for different $\Lambda$
they are orthogonal.  Thus the orthonormality or unitarity
and completeness relations for the Clebsch-Gordan coefficients
are:
     \begin{eqnarray}
     \sum\limits_{M_1,M_2}\;
     C^{J_1}_{M_1}\;^{J_2}_{M_2}\;^{J_3\Lambda}_{M_3}\;
     {C^{J_1}_{M_1}\;^{J_2}_{M_2}\;^{J^{\prime}_3\Lambda^{\prime}}
     _{M_3^{\prime}}}^* &=&
     \delta_{\Lambda^{\prime}\Lambda}\;
     \delta_{J^{\prime}_3 J_3}\;\delta_{M^{\prime}_3 M_3}\;,
     \nonumber\\
     \sum\limits_{\Lambda J_3,M_3}\;
     C^{J_1}_{M_1}\;^{J_2}_{M_2}\;^{J_3\Lambda}_{M_3}\;
     {C^{J_1}_{M_1^{\prime}}\;^{J_2}_{M_2^{\prime}}\;
     ^{J_3\Lambda}_{M_3}}^* &=&
     \delta_{M^{\prime}_1 M_1}\;\delta_{M^{\prime}_2 M_2}\;.
     \end{eqnarray}

\noindent
The statement that for each 
$\Lambda, \Psi^{(J_{3}\Lambda)}_{M_{3}}$ transforms according to the UIR ${\cal D}^{(J_3)}$ of $G$ leads to:
     \begin{eqnarray}
     \sum\limits_{M_1,M_2}\;
     C^{J_{1}}_{M_{1}}\;^{J_{2}}_{M_{2}}\;^{J_{3}\Lambda}_{M_{3}}\;
     {\cal D}^{(J_1)}_{M^{\prime}_{1} M_{1}}(g)\;
     {\cal D}^{(J_2)}_{M^{\prime}_{2} M_{2}} (g) 
     =  \sum\limits_{M^{\prime}_{3}}\; 
      C^{J_1}_{M^{\prime}_1}\;^{J_{2}}_{M^{\prime}_2}\;
      ^{J_3\Lambda}_{M_{3}^{\prime}} \;
      {\cal D}^{(J_{3})}_{M^{\prime}_{3}\; M_{3}}(g),       
      \end{eqnarray}

\noindent
from which follows, using (A.5), the result for the 
product of any two ${\cal D}$-matrices:
     \begin{eqnarray}
     {\cal D}^{(J_1)}_{M^{\prime}_{1} M_{1}}(g)\;
     {\cal D}^{(J_2)}_{M^{\prime}_{2} M_{2}}(g)=
     \sum\limits_{\Lambda J_3 M_3^{\prime} M_3}\;
     C^{J_{1}}_{M_{1}^{\prime}}\;
     ^{J_{2}}_{M_{2}^{\prime}}\;
     ^{J_{3}\Lambda}_{M_{3}^{\prime}}\;
     {C^{J_{1}}_{M_{1}}\; ^{J_{2}}_{M_{2}}\;
     ^{J_{3}\Lambda}_{M_{3}}}^*\;
     {\cal D}^{(J_3)}_{M^{\prime}_{3} M_{3}}(g)\,.
     \end{eqnarray}

Lastly we consider the Wigner-Eckart theorem, and the definition 
and properties of unit tensor operators within a UIR.  A tensor
operator of type $J_2$ connecting the two UIR's $J_1$ and
$J_3$ is a collection of operators
     \begin{eqnarray}
     T^{J_2}_{M_2}: {\cal H}^{(J_1)}\longrightarrow
     {\cal H}^{(J_3)}\;     ,
     \end{eqnarray}

\noindent
obeying the transformation rule
     \begin{eqnarray}
     {\cal D}^{(J_3)}(g)\; T^{J_2}_{M_2}
     {\cal D}^{(J_1)}(g)^{-1} =
     \sum\limits_{M^{\prime}_2} 
     {\cal D}^{(J_2)}_{M^{\prime}_{2} M_{2}}(g)\;
     T^{J_2}_{M^{\prime}_2}\;.
     \end{eqnarray}

\noindent
The matrix elements of such a set of operators between the two
sets of basis states involves a collection of reduced matrix
elements labelled by the Clebsch-Gordan multiplicity label
$\Lambda$ and accompanied by corresponding Clebsch-Gordan
coefficients:
     \begin{eqnarray}
     \left(\Psi^{(J_3)}_{M_3},\;T^{J_2}_{M_2}\;
     \Psi^{(J_1)}_{M_1}\right) = \sum\limits_{\Lambda}\;
     {C^{J_{1}}_{M_{1}}\;^{J_{2}}_{M_{2}}\;^{J_{3}\Lambda}
     _{M_{3}}}^*\; \langle  J_3\|T^{J_2}\|J_1 \rangle _{\Lambda}\; .
     \end{eqnarray}

\noindent
The occurrence of the complex conjugate of the Clebsch-Gordan
coefficients is to be noted.  One can then express $T^{J_2}_{M_2}$
explicitly as:
      \begin{eqnarray}
      T^{J_2}_{M_2} =\sum\limits_{\Lambda M_1 M_3}\;
      {C^{J_{1}}_{M_{1}}\;^{J_{2}}_{M_{2}}\;^{J_{3}\Lambda}
      _{M_{3}}}^*\;\langle J_3\|T^{J_2}\|J_1 \rangle_{\Lambda}\;
      \Psi^{(J_3)}_{M_3}\;\Psi^{(J_1)^{\dag}}_{M_1}\,.
       \end{eqnarray}     

Within a given UIR ${\cal D}^{(J_0)}$ of $G$ on 
${\cal H}^{(J_0)}$, eqn.(A.11) leads to the definition of a
complete set of unit tensor operators $U^{J\Lambda}_M$ as
follows:
     \begin{eqnarray}
     U^{J\Lambda}_M =\sum\limits_{M_0 M^{\prime}_0}\;
     {C^{J_0}_{M_0}\;^{J}_{M}\;^{J_0\Lambda}_{M^{\prime}_0}}^*
     \;\Psi^{(J_0)}_{M_0}\;{\Psi^{(J_0)}_{M^{\prime}_0}}^{\dag}\,,
     \end{eqnarray}

\noindent
where we have chosen specially simple values for the reduced
matrix elements.  These unit tensors obey, as a
particular case of (A.9):
     \begin{eqnarray}
     {\cal D}^{(J_0)} (g)\; U^{J\Lambda}_M\;
     {{\cal D}^{(J_0)}(g)}^{-1} =
     \sum\limits_{M^{\prime}}\;{\cal D}^{(J)}_{M^{\prime}M}
     (g)\; U^{J\Lambda}_{M^{\prime}}\,.
     \end{eqnarray}

\noindent
One can also easily establish their trace orthogonality:
     \begin{eqnarray}
     \mbox{Tr}\left({U^{J^{\prime}\Lambda^{\prime}}_{M^{\prime}}}^{\dag}
     \;U^{J\Lambda}_M\right) =
     \displayfrac{N_{J_{0}}}{N_{J}}\;\delta_{\Lambda^{\prime}
      \Lambda}\;\delta_{J^{\prime}J}\;\delta_{M^{\prime}M}\,.
     \end{eqnarray}

\noindent
Therefore any operator $A$ on ${\cal H}^{(J_0)}$ is uniquely
expressible in the form
     \begin{eqnarray}
     A &=& \sum\limits_{\Lambda J M}\;
     a^{J\Lambda}_M\;U^{J\Lambda}_M \;,\nonumber\\
     a^{J\Lambda}_M&=& \displayfrac{N_J}{N_{J_{0}}}\;
     \mbox{Tr}\left({U^{J\Lambda}_M}^{\dag} A\right)\,.
     \end{eqnarray}

\noindent
In Section 3 we have used such formulae in a basis
adapted to $H$.\\

\def\theequation{B.\arabic{equation}}
\appendix{\bf{Appendix B}: Induced representations on coset spaces
and reciprocity theorems\\}
\setcounter{equation}{0}

Here we outline the construction of induced UR's of $G$ 
starting from UIR's of $H$, and the reciprocity theorem 
which tells us in detail the irreducible contents of such
UR's of $G$.  A direct construction of a class of UR's
of a semidirect product of $G$ by a certain abelian group
(similar to the Euclidean and Poincar\'{e} groups) proves
practically useful in this context.\\

\noindent
{\bf The inducing construction }\\

  The UIR $D^{(j)}(h)$ of
$H$ is defined on the Hilbert space ${\cal H}^{(j)}$ of dimension
$N_j$.  Consider functions $\phi : G\rightarrow {\cal H}^{(j)}$ 
satisfying the following (right) covariance law under $H$:
     \begin{eqnarray}
      g\,\in\,G\rightarrow \phi(g)&\in & 
      {\cal H}^{(j)} \,,\nonumber\\
      \phi(gh)&=& D^{(j)}(h^{-1})^* \;\phi(g),\nonumber\\
      \mbox{ie},\;\; \;\phi_m(gh)&=&\sum\limits_{m^{\prime}}
      \;D^{(j)}_{m^{\prime}m}(h) \;\phi_{m^{\prime}}
      (g)\,.
       \end{eqnarray}

\noindent
(We avoid using letters $\psi,\Psi$ for these vector valued 
functions  on $G$ since they have been used in the main 
text with specific meanings).  We now define an (left)
action by $G$ on such $\phi$:
     \begin{eqnarray}
     ({\cal U}(g)\phi)(g^{\prime}) =\phi(g^{-1}g^{\prime})\,.
     \end{eqnarray}

\noindent
The representation property is obvious, and so also the 
compatibility  of the condition (B.1) and the action (B.2),
ie., the latter respects the former.  Let $\Sigma=G/H$ be
(as in the text) the space of right cosets in $G$ with
respect to $H$, and let $\ell(q)$ be a choice of (local) coset 
representatives $\Sigma\rightarrow G$.  Then it is clear that 
the independent information in a $\phi$ obeying (B.1) is 
contained in its values at coset representatives:
     \begin{eqnarray}
     q\,\in\,\Sigma:\;\;\; \phi_0(q) = \phi(\ell(q))\,.
     \end{eqnarray}

\noindent
On these the action by $G$ is easily computed:
     \begin{eqnarray}
     {\cal U}(g) \phi= \phi^{\prime}:\;\;&& \nonumber\\
     \phi^{\prime}_0(q)&=& \phi^{\prime}(\ell(q))\nonumber\\
     &=&\phi(g^{-1}\ell(q))\nonumber\\
     &=& \phi(\ell(g^{-1}q)\;
     \ell(g^{-1}q)^{-1}\;g^{-1}\ell(q))\nonumber\\
     &=&D^{(j)}\left(\ell(q)^{-1} g\;\ell(g^{-1}q)\right)^*\;
     \phi_0(g^{-1}q)\; ,\nonumber\\
     \mbox{ie}\,,\;\;\;\phi^{\prime}_{0,m}(q) &=&
     \sum\limits_{m^{\prime}}\;D^{(j)}_{m^{\prime}m}
     \;(\ell(g^{-1}q)^{-1}\;g^{-1}\ell(q))\;
     \phi_{0,m^{\prime}}(g^{-1}q)\; .
     \end{eqnarray}

\noindent
We can now formally define the Hilbert space for these 
\lq wave functions \rq, in such a way that the operators 
${\cal U}(g)$ are unitary.  We use the following notation:
     \begin{eqnarray}
     L^2(\Sigma, {\cal H}^{(j)}) =
     \left\{\phi_0(q)\,\in\,{\cal H}^{(j)}\;\bigg|\;
     q\,\in\,\Sigma, \;\|\phi_0\|^2 
      = \int\limits_{\Sigma} d\mu(q) (\phi_0(q),\, \phi_0(q))
     _{{\cal H}^{(j)}} < \infty\right\}\,.
     \end{eqnarray}

\noindent
Here $d\mu(q)$ is the $G$-invariant normalised volume 
element on $\Sigma$, and it is obvious that unitarity 
of $D^{(j)}$ leads to unitarity of ${\cal U}(g)$.
This UR of $G$ is said to be induced from the UIR 
$D^{(j)}$ of $H$, and we will denote it as 
${\cal D}^{(\mbox{ind},j)}$ (the dependence on $H$ being 
left implicit).  Combining eqns (B.4,5) we see that we can 
introduce an (ideal) orthonormal basis 
$|q,m\rangle$ for $L^2(\Sigma,{\cal H}^{(j)})$ with these 
properties:   
     \begin{eqnarray}
     \phi_{o,m}(q) &=& \langle q,m|\phi_0\rangle\,,\nonumber\\
     \langle q^{\prime},m^{\prime}|q,m \rangle &=& \delta(q^{\prime},
     q) \delta _{m^{\prime}m} \;;\nonumber\\
     {\cal U}(g)|q,m \rangle&=& \sum\limits_{m^{\prime}}\;
     {\cal D}^{(j)}_{m m^{\prime}}(\ell(q)^{-1}
     g^{-1}\ell(gq))|gq,m^{\prime} \rangle\,.
     \end{eqnarray}

\noindent
This can be viewed as a standard Wigner form for the UR
${\cal D}^{(\mbox{ind},j)}$ of $G$.

Now the main question is: how often does the UIR
${\cal D}^{(J)}$ of $G$ occur in the UR 
${\cal D}^{(\mbox{ind},j)}$ of $G$, and in case there is
nontrivial multiplicity is there a natural way to
choose a multiplicity label in an orthonormal manner?
To answer these, we turn to a convenient construction of a 
`Master UR' of a certain semidirect product group ${\cal G}$
involving $G$, originally studied in the context of
strong coupling theory\cite{CGS,NM}.\\

\noindent
{\bf The group ${\cal G}$ and the $CGS$ construction}\\

Choose some UIR ${\cal D}^{(J_{0})}$ of $G$ (obeying a 
condition to be given later) and consider a group ${\cal G}$ 
defined as the semidirect product of $G$ by an abelian 
part $P^{(J_{0})}$ whose generators belong to 
${\cal D}^{(J_{0})}$.  It is convenient to express
the structure of ${\cal G}$ partly in finite form
(the $G$ part) and partly in terms of infinitesimal
generators (the abelian part).  Thus we look for unitary
operators $\overline{U}(g), g\,\in\,G$, and additional
(possibly nonhermitian) operators $P^{(J_{0})}_{M_{0}}$
obeying the relations
     \begin{mathletters}
     \begin{eqnarray}
     \overline{U}(g^{\prime})\overline{U}(g)&=&
      \overline{U}(g^{\prime}g) \;;\\
      \overline{U}(g)^{-1} P^{(J_{0})}_{M_{0}}
      \overline{U}(g) &=&\sum\limits_{M^{\prime}_{0}}
      {\cal D}^{(J_{0})}_{M_{0}M_{0}^{\prime}}(g)
      P^{(J_{0})}_{M_{0}^{\prime}}\; ;\\
      \protect\left[P^{(J_{0})}_{M_{0}}\,,\, 
      P^{(J_{0})}_{M_{0}^{\prime}}\right. &\mbox{or} &
      \left.{P^{(J_{0})}_{M_{0}^{\prime}}}^{\dag}
      \protect\right] = 0\;.
      \end{eqnarray}
       \end{mathletters}

\noindent
These relations define ${\cal G}$, and the analogy to the
structures of $E(3)$ or the Poincar\"{e} group is evident;
therefore we can refer to the $P^{(J_{0})}_{M_{0}}$ as
`momenta'.

We now set up a solution to these relations on the space
${\cal H}^{(\mbox{reg})}=L^2(G,{\cal C})$ of the regular
representation ${\cal D}^{(\mbox{reg})}$ of $G$.  We introduce 
ideal basis vectors $|g\rangle$ obeying
     \begin{eqnarray}
     \langle g^{\prime}|g\rangle = \delta(g^{-1}g^{\prime})\,.
      \end{eqnarray}

\noindent
Choose now some numerical (possibly complex) values 
$p^{(J_{0})}_{M_{0}}$ as possible eigenvalues of the
$P^{(J_0)}_{M_0}$, and define $\overline{U}(g), P^{(J_0)}
_{M_0}$ on the basis kets 
$|g\rangle\,\in\,{\cal H}^{(\mbox{reg})}$ by:
     \begin{eqnarray}
     \overline{U}(g) |g^{\prime} \rangle &=& |gg^{\prime} \rangle\;,
     \nonumber\\
     P^{(J_{0})}_{M_{0}} |g^{\prime} \rangle &=&
     ({\cal D}^{(J_{0})}(g^{\prime}) p^{(J_{0})})_{M_{0}}
      |g^{\prime} \rangle\;.
     \end{eqnarray}

\noindent
One can verify that $\overline{U}(g)$ are unitary, and that 
all the relations (B.7) are obeyed, so we have here a certain
master UR of ${\cal G}$ uniquely specified by the choice of
$p^{(J_{0})}$.  The basis $|g\rangle$ is one in which the `momenta'
are all simultaneously diagonal, and this is the essence
of the CGS construction.

This UR of ${\cal G}$ can be analysed in two interesting
ways by using two separate bases for ${\cal H}^{(\mbox{reg})}$.
On the one hand we can exploit the orthogonality and 
completeness of the UIR's of $G$ as expressed by eqn.(A.2), 
and so introduce a basis $|JMN\rangle$ defined and behaving as 
follows:
     \begin{eqnarray*}
     |JMN \rangle = N^{1/2}_J \int\limits_{G} dg\;
     {\cal D}^{(J)}_{MN} (g) | g \rangle\; ,
     \end{eqnarray*}
           \begin{mathletters}
      \begin{eqnarray}
     \langle J^{\prime}M^{\prime}N^{\prime} |JMN\rangle&=&
     \delta_{J^{\prime}J}\delta_{M^{\prime}M}
     \delta_{N^{\prime}N}\; ; \\
     \overline{U}(g) |JMN \rangle &=&\sum\limits_{M^{\prime}}
     {\cal D}^{(J)}_{M^{\prime}M}(g)^* |
     JM^{\prime}N \rangle \;.
     \end{eqnarray}
     \end{mathletters}

\noindent
In this basis in which the regular representation of $G$ is 
fully reduced, we can exploit the information given in 
Appendix A to show that the matrix elements of the momenta 
$P^{(J_{0})}_{M_{0}}$ have the following form:
     \begin{eqnarray}
     \langle J^{\prime}M^{\prime}N^{\prime}|\;
     P^{(J_{0})}_{M_{0}}\;
     |JMN\rangle =
     \sqrt{\displayfrac{N_{J}}{N_{J^{\prime}}}}
     \;\sum\limits_{\Lambda N_{0}}\;p^{(J_{0})}_{N_{0}}
      C^{J}_{M}\;^{J_{0}}_{M_{0}}\;
     ^{J^{\prime}\Lambda}_{M^{\prime}}\;
     {C^{J}_{N}\;^{J_{0}}_{N_{0}}\;
     ^{J^{\prime}\Lambda}_{N^{\prime}}}^*\;.
     \end{eqnarray}

\noindent
This means that the reduced matrix element of
$P^{(J_{0})}$ with multiplicity label $\Lambda$ is 
(see eqn.(A.10))).
     \begin{eqnarray}
      \langle  J^{\prime}N^{\prime}\|P^{(J_0)}\|JN \rangle _{\Lambda} =
      \sqrt{\displayfrac{N_J}{N_{J^{\prime}}}}\;\sum\limits
      _{N_{0}}\;p^{(J_0)}_{N_0}\;
      {C^{J}_{N}\;^{J_{0}}_{N_{0}}\;
     ^{J^{\prime}\Lambda}_{N^{\prime}}}^*\,.
      \end{eqnarray}

\noindent
We will use this in a moment.

The other way to exploit the CGS construction (B.9) is to pass 
to a description in terms of a coset space.  At this 
point we assume that the stability group of the numerical
momentum $p^{(J_0)}_{M_0}$ is the subgroup $H\subset G$:
     \begin{eqnarray}
     h\,\in \,H:\;\;\; {\cal D}^{(J_0)} (h)\;
     p^{(J_0)} = p^{(J_0)} \,.
     \end{eqnarray}

\noindent
Thus the condition on the choice of the UIR $J_0$ of $G$
while constructing ${\cal G}$ is that ${\cal H}^{(J_0)}$
must contain (at least) one $H$-scalar state.  We 
then express a general $g^{\prime}\,\in\,G$ as the
product $g^{\prime}=\ell(q)h$ of a coset representative and
a subgroup element:
     \begin{eqnarray}
     |g^{\prime} \rangle  &=& |q, h) \,,\nonumber\\
     (q^{\prime},h^{\prime}|q,h)&=&
     \delta(q^{\prime},q)\;\delta(h^{-1}h^{\prime})\,.
     \end{eqnarray}

\noindent
Then eqn.(B.9) appear thus:
     \begin{eqnarray}
     \overline{U}(g)|q,h) &=& |g \ell(q)h \rangle \nonumber\\
     &=&|\ell(gq) \ell(gq)^{-1} g\ell(q)h \rangle \nonumber\\
     &=&|gq, \ell(gq)^{-1} g\ell(q)h)\;;\nonumber\\
     P^{(J_0)}_{M_0} |g,h)&=&\left({\cal D}^{(j_0)}(\ell
     (q)) p^{(J_0)}\right)_{M_0} |q,h)\;.
     \end{eqnarray}

\noindent
The key point is that in the last relation the eigenvalues
of the momenta are independent of $h$, precisely because
of eqn.(B.13)..  To arrive at basic states behaving in the
Wigner form (B.6) under $\overline{U}(g)$ we just have to
exploit the regular representation of $H$ in the same way as
we did for $G$ in eqn.(B.10).  So from $|q,h)$ we pass to
$|q,jmn\rangle $:
     \begin{eqnarray}
     |q,jmn\rangle &=&N^{1/2}_j \int\limits_{H} dh\;D^{(j)}_{mn}(h)
     |q,h) \;,\nonumber\\
     \langle q^{\prime},j^{\prime}m^{\prime}n^{\prime}|q,jmn\rangle  &=&
     \delta(q^{\prime},q) \delta_{j^{\prime}j} 
     \delta_{m^{\prime}m} \delta_{n^{\prime}n}\;.
     \end{eqnarray}

\noindent
In this basis we find:
     \begin{eqnarray}
     \overline{U}(g) |q,jmn\rangle &=&\sum\limits_{m^{\prime}}\;
     D^{(j)}_{mm^{\prime}} (\ell(q)^{-1}g^{-1}\ell(gq))|
     gq, jm^{\prime}n \rangle  \;,\nonumber\\
     P^{(J_0)}_{M_0} |q,jmn\rangle &=&\left({\cal D}^{(j_0)}(\ell
     (q)) p^{(J_0)}\right)_{M_0} |q,jmn \rangle \; .
     \end{eqnarray}

\noindent
All the operators of ${\cal G}$, both $\overline{U}(g)$ and 
$P^{(J_0)}_{M_0}$, conserve the quantum numbers $j$ and $n$.
So if these are kept fixed, and only $q$ and $m$ are allowed
to vary, we see that we have exactly recovered eqn.(B.6).
This shows that the CGS UR of ${\cal G}$ corresponding to a
$p^{(J_0)}_{M_0}$ with stability group $H\subset G$ contains each 
induced UR ${\cal D}^{(\mbox{ind},j)}$ of $G$ exactly $N_j$ times.

On the other hand, we can link up now to the results (B.10)
in the basis $|JMN\rangle $ by adapting the choice of labels
$M,N,\ldots$ to reduction with respect to the subgroup
$H$.  As described in Appendix A, this makes $M,N\ldots$ into
triples $\mu k m, \nu j n,\ldots$, and then eqn.(B.10,12) 
become:
     \begin{eqnarray}
     \overline{U}(g)|J\;\mu k m\;\nu j n\rangle &=&
     \sum\limits_{\mu^{\prime}k^{\prime}m^{\prime}}\;
     {\cal D}^{(J)}_{\mu^{\prime}k^{\prime}m^{\prime},
     \mu k m}(g)^* |J\;\mu^{\prime}k^{\prime}m^{\prime}\;
     \nu j n \rangle  \;,\nonumber\\
     \langle J^{\prime}\;\nu^{\prime}j^{\prime}n^{\prime}\|
     P^{(J_0)}\|J\;\nu j n\rangle &=&
     \delta_{j^{\prime}j} \delta_{n^{\prime}n}
     \sqrt{\displayfrac{N_J}{N_{J^{\prime}}}}
      \sum\limits_{\nu_{0}}\;p^{(J_0)}_{\nu_{0} 00}\;
     {C^{J}_{\nu j n}\;^{J_0}_{\nu_{0} 00}\;
     ^{J^{\prime}\Lambda}_{\nu^{\prime}j n}}^*    \; .
     \end{eqnarray}

\noindent
There are as many independent components to $p^{(J_0)}$ as 
there are $H$-scalar states in ${\cal D}^{(J_0)}$.  So while
$\overline{U}(g)$ conserve $\nu j n, P^{(J_0)}_{M_0}$
conserve only $j$ and $n$, but not the multiplicity labels
$\nu^{\prime}$ and $\nu$.  Realising that from the original
basis $|g\rangle $ for ${\cal H}^{(\mbox{reg})}$ we have arrived
in two ways, via the sequences $|g\rangle \rightarrow |JMN\rangle \rightarrow
|J\;\mu k m\; \nu jn\rangle $ and $|g\rangle \rightarrow|q,h)\rightarrow
|q,jmn\rangle $, at two alternative bases for the same UR
of ${\cal G}$, in which the actions by $\overline{U}(g)$ and
$P^{(J_0)}_{M_0}$ are respectively given by eqn.(B.18) and
eqn.(B.17), we come to the following conclusions:
     \begin{eqnarray}
     &&Sp(|J\;\mu k m\;\nu j n\rangle |J\;\mu  k m \nu\;\mbox{varying},
     \;j n\;\mbox{fixed})\nonumber\\
     &&~~~~~~~~~~~~~~~~ =Sp(|q, jmn\rangle  |q, m\;\mbox{varying},\;jn\;
     \mbox{fixed})\, ,
     \end{eqnarray}

\noindent
and the corresponding subspace of ${\cal H}^{(\mbox{reg})}$
carries exactly once the induced UR ${\cal D}^{(\mbox{ind},j)}$ of
$G$.  Comparing this with the reduced matrix element result
(B.18) we then see that this UR of $G$ contains the UIR
${\cal D}^{(J)}$ of $G$ as often as ${\cal D}^{(J)}$ 
contains the UIR $D^{(j)}$ of $H$, which is the reciprocity
theorem; the index $\nu$ catalogues (in an orthonormal
way) these several occurrences of ${\cal D}^{(J)}$.

We appreciate that in the final statement of the reciprocity
theorem the representation ${\cal D}^{(J_0)}$ and the momenta
$P^{(J_0)}_{M_0}$ have disappeared; they play only an 
intermediate role in the CGS construction and in recognising 
that we have two equally good bases for the Hilbert space 
carrying the UR ${\cal D}^{(\mbox{ind},j)}$ of $G$.\\

\noindent
{\bf ACKNOWLEDGEMENTS:} NM thanks Profs. G. Marmo and G. Morandi at the
Departments of Physics of the Universities of Napoli and Bologna respectively
for helpful discussions and hospitality during visits when this work was
initiated.


\end{document}